\documentclass{emulateapj}

\usepackage{amssymb}
\usepackage{psfig}
\usepackage{graphics,graphicx}
\usepackage{subfigure}
\usepackage{epsfig}
\usepackage{natbib}
\usepackage{longtable}
\usepackage{lscape}

\begin{document}


\title{Bright broad-band afterglows of gravitational
wave bursts from mergers of binary neutron stars}


\author{He Gao\altaffilmark{1,2}, Xuan Ding\altaffilmark{1},
Xue-Feng Wu\altaffilmark{1,6,7,*}, Bing Zhang\altaffilmark{3,4,2,*},
Zi-Gao Dai\altaffilmark{5,*}}


\altaffiltext{1}{Purple Mountain Observatory, Chinese Academy of
Sciences, Nanjing, 210008, China} \altaffiltext{2}{Department of
Physics and Astronomy, University of Nevada Las Vegas, NV 89154,
USA} \altaffiltext{3}{Department of Astronomy, Peking University,
Beijing 100871, China} \altaffiltext{4}{Kavli Institute of Astronomy
and Astrophysics, Peking University, Beijing 100871, China}
\altaffiltext{5}{School of Astronomy and Space Science, Nanjing
University, Nanjing 210093, China }\altaffiltext{6}{Chinese Center
for Antarctic Astronomy, Chinese Academy of Sciences, Nanjing,
210008, China}\altaffiltext{7}{Joint Center for Particle Nuclear
Physics and Cosmology of Purple Mountain Observatory-Nanjing
University, Chinese Academy of Sciences, Nanjing 210008, China}
\altaffiltext{*}{xfwu@pmo.ac.cn;~~zhang@physics.unlv.edu;~~dzg@nju.edu.cn}

\begin{abstract}
If double neutron star mergers leave behind a massive magnetar
rather than a black hole, a bright early afterglow can follow the
gravitational wave burst (GWB) even if there is no short gamma-ray
burst (SGRB) - GWB association or there is an association but the
SGRB does not beam towards earth. Besides directly dissipating the
proto-magnetar wind as suggested by Zhang, we here suggest that the
magnetar wind could push the ejecta launched during the merger
process, and under certain conditions, would reach a relativistic
speed. Such a magnetar-powered ejecta, when interacting with the
ambient medium, would develop a bright broad-band afterglow due to
synchrotron radiation. We study this physical scenario in detail,
and present the predicted X-ray, optical and radio light curves for
a range of magnetar and ejecta parameters. We show that the X-ray
and optical lightcurves usually peak around the magnetar spindown
time scale ($\sim 10^3-10^5$ s), reaching brightness readily
detectable by wide-field X-ray and optical telescopes, and remain
detectable for an extended period. The radio afterglow peaks later,
but is much brighter than the case without a magnetar energy
injection. Therefore, such bright broad-band afterglows, if detected
and combined with GWBs in the future, would be a probe of massive
millisecond magnetars and stiff equation-of-state for nuclear
matter.
\end{abstract}





\section{Introduction}

The next generation gravitational-wave detectors, such as Advanced
LIGO \citep{ligo}, Advanced VIRGO \citep{virgo} and KAGRA
\citep{kuroda10} interferometers, are expected to detect GW signals
from mergers of two compact objects. These gravitational wave bursts
(GWBs) have well defined ``chirp'' signal, which can be
unambiguously identified. Once detected, the GW signals would open a
brand new channel for us to study the universe, especially the
physics in the strong field regime. Due to the faint nature of GWs,
an associated electromagnetic (EM) emission signal in coincidence
with a GWB in both trigger time and direction would increase the
signal-to-noise ratio of the GW signal, and therefore would be
essential for its identification.

One of the top candidates of GWBs is merger of two neutron stars
(i.e. NS-NS mergers) \citep{taylor82,kramer06}. The EM signals
associated with such an event include a short gamma-ray burst (SGRB)
\citep{eichler89,rosswog12,gehrels05,barthelmy05a,berger11}, an
optical ``macronova'' \citep{lipaczynski98,kulkarni05,metzger10},
and a long lasting radio afterglow
\citep{nakar11,metzger12,piran12}. Numerical simulations show that
binary neutron star mergers could eject a fraction of the materials,
forming a mildly anisotropic outflow with a typical velocity about
$0.1-0.3 c$ (where $c$ is the speed of light), and a typical mass
about $10^{-4} \sim 10^{-2}M_{\odot}$
\citep[e.g.][]{rezzolla11,rosswog12,hotokezaka12}. The radioactivity
of this ejecta powers the macronova and the interaction between the
ejecta and the ambient medium is the source of radio afterglow.
Usually, the merger product is assumed to be a black hole or a
temporal hyper-massive neutron star which survives 10-100 ms before
collapsing into the black hole
\citep[e.g.][]{rosswog03,aloy05,shibata05,rezzolla11,rosswog12}.
Nonetheless, recent observations of Galactic neutron stars and NS-NS
binaries suggest that the maximum NS mass can be high, which is
close to the total mass of the NS-NS systems \citep[][and references
therein]{dai06,zhang13}. Indeed, for the measured parameters of 6
known Galactic NS binaries and a range of equations of state, the
majority of mergers of the known binaries will form a massive
millisecond pulsar and survive for an extended period of time
\citep{morrison04}. When the equation of state of nuclear matter is
stiff (see arguments in \cite{dai06} and \cite{zhang13} and
references therein), a stable massive neutron star would form after
the merger. This newborn massive neutron star would be
differentially rotating. The dynamo mechanism may operate and
generate an ultra-strong magnetic field
\citep{duncan92,kluzniak98,dailu98a}, so that the product is very
likely a millisecond magnetar. Evidence of a magnetar following some
SGRBs has been collected in the Swift data
\citep{rowlinson10,rowlinson12}, and magnetic activities of such a
post-merger massive neutron star have been suggested to interpret
several X-ray flares and plateau phase in SGRBs
\citep{dai06,gao06,fanxu06}.

Since both the gravitational wave signal and the millisecond magnetar
wind both nearly isotropic, a bright electromagnetic signal can be
associated with a NS-NS merger GWB regardless of
whether there is a short gamma-ray burst (SGRB) - GWB association
\citep{zhang13}. Even if there is an association, most GWBs would not
be associated with the SGRB since SGRBs are collimated.  \cite{zhang13}
proposed that the near-isotropic magnetar wind of a post-merger millisecond
magnetar would undergo magnetic dissipation \citep{zhangyan11} and power a
bright X-ray afterglow emission. Here we suggest that after partially
dissipating the magnetic energy,
a significant fraction $\xi$ of the magnetar spin energy would be used
to push the ejecta, which drives a strong forward shock into the
ambient medium. The continuous injection of the Poynting flux into
the blast wave modifies the blast wave dynamics and leads to rich
radiation signatures \citep{dailu98b,zhangmeszaros01a,dai04}. Figure
1 presents a physical picture for several EM emission components
appearing after the merger. We here study the dynamics of such an
interaction in detail, and calculate broadband afterglow emission
from this forward shock.

\begin{figure}[h!!!]
\centerline{\psfig{file=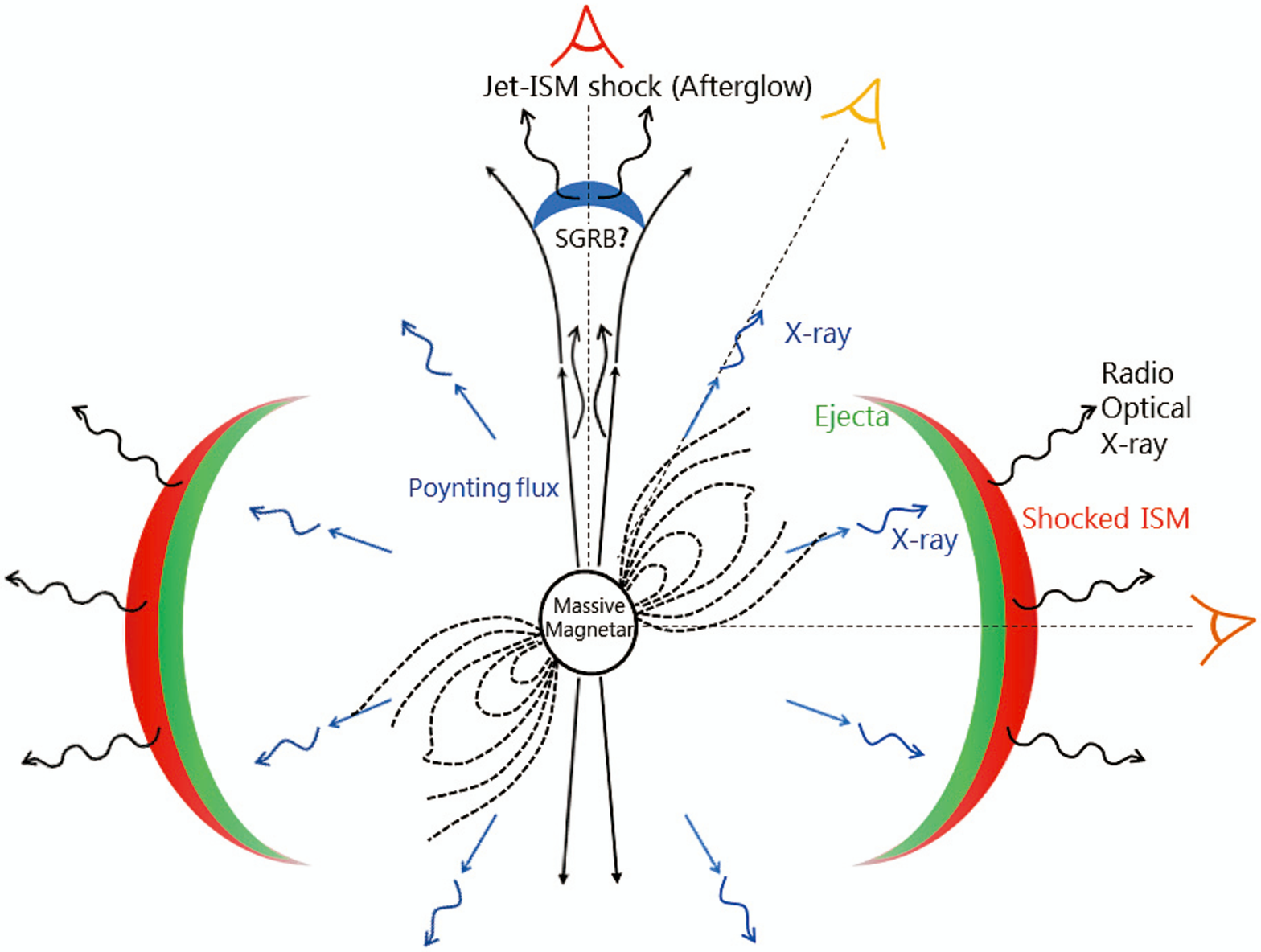,width=3in,angle=-90}} \caption{A
physical picture for several EM emission components appearing after
the merger. A massive millisecond magnetar is formed at the central
engine. Near the spin axis, there might be a SGRB jet. An observer
towards this jet (red observer) would see a SGRB. At larger angles
(yellow observer), a free magnetar wind may be released, whose
dissipation would power a bright X-ray afterglow \citep{zhang13}. At
even larger angles (orange observer), the magnetar wind is confined
by the ejecta (green shell). The interaction between the magnetar
wind and ejecta would trigger magnetic dissipation of the wind and
also power a bright X-ray afterglow \citep{zhang13}. After releasing
some dissipated energy, a significant fraction of the spinning
energy would push the ejecta and shock into the ambient medium
\citep{dailu98b,zhangmeszaros01a}. Synchrotron emission from the
shocked medium (red shell) would power brighter X-ray, optical and
radio afterglow emission, which is calculated in this work.}
\label{cartoon}
\end{figure}

\section{The model}

The postmerger hyper-massive neutron star may be near the break up
limit, so that the total spin energy $E_{\rm{rot}}=(1/2)I
\Omega_{0}^{2} \simeq 2\times 10^{52} I_{45} P_{0,-3}^{-2} ~{\rm
erg}$ (with $I_{45} \sim 1.5$ for a massive neutron star) may be
universal. Here $P_{0} \sim 1$ ms is the initial spin period of the
proto-magnetar. Throughout the paper, the convention $Q=10^n Q_n$ is
used for cgs units, except for the ejecta mass $M_{\rm ej}$, which
is in unit of solar mass $M_{\odot}$. Given nearly the same total
energy, the spin-down luminosity and the characteristic spin down
time scale critically depend on the polar-cap dipole magnetic field
strength $B_{p}$ \citep{zhangmeszaros01a}, i.e. $L_{\rm sd} = L_{\rm
sd,0}/(1+t/T_{\rm sd})^2$, where $L_{\rm sd,0} \simeq 10^{49} ~{\rm
erg~s^{-1}}~B^{2}_{p,15}R_{6}^{6}P_{0,-3}^{-4}$, and the spin down
time scale $T_{\rm{sd}} \simeq 2 \times 10^3 ~{\rm s}~I_{45}
B_{p,15}^{-2} P_{0,-3}^2 R_6^{-6} \simeq E_{\rm{rot}}/L_{\rm sd,0}$,
where $R=10^6R_6$ cm is the stellar radius\footnote{In principle,
besides dipole spindown, the proto-magnetar may also lose spin
energy via gravitational radiation \citep{zhangmeszaros01a,corsi09,fan13}.
This effect is neglected in the following modeling for simplicity.}.

After the internal dissipation of the magnetar wind that powers the
early X-ray afterglow \citep{zhang13}, the remaining spin
energy would be added to the blastwave. The dynamics of the
blastwave depends on the magnetization parameter $\sigma$ of the
magnetar wind after the internal dissipation. Since for the confined
wind, magnetic dissipation occurs upon interaction between the wind
and the ejecta, in this paper, we assume that the wind is still
magnetized (moderately high $\sigma$), so that there is no strong
reverse shock into the magnetar wind
\citep{zhangkobayashi05,mimica09}\footnote{If, on the other hand,
the wind is already leptonic matter dominated, a reverse shock can
be developed, which would predict additional radiation signatures
\citep{dai04}.}. As a result, the remaining spin energy is
continuously injected into the blastwave with a luminosity $L_0 =
\xi L_{\rm sd,0}$, where $\xi < 1$ denotes the fraction of the spin
down luminosity that is added to the blastwave. The evolution of the
blastwave can be described by a system with continuous energy
injection \citep{dailu98b,zhangmeszaros01a}.

The newly formed massive magnetar is initially hot. A Poynting flux
dominated outflow is launched $\sim 10$ s later, when the
neutrino-driven wind is clean enough \citep{metzger11}. At this
time, the front of the ejecta traveled a distance $\sim 6 \times
10^{10}$ cm (for $v \sim 0.2 c$), with a width $\Delta \sim 10^7$
cm. The ultra-relativistic magnetar wind takes $\sim 2$ s to catch
up the ejecta, and drives a forward shock into the ejecta. Balancing
the magnetic pressure and the ram pressure of shocked fluid in the
ejecta, one can estimate the shocked fluid speed as $v_s \sim
10^{-4} c L_{0,47}^{1/2} \Delta_7^{1/2} M_{\rm ej,-3}^{-1/2}$, which
is in the same order of forward shock speed. So the forward shock
would cross the ejecta in around $t_\Delta \sim \Delta / v_s \sim
3~{\rm s}~ L_{0,47}^{-1/2} \Delta_7^{1/2} M_{\rm ej,-3}^{1/2}$. Note
that when calculating magnetic pressure, we have assumed a toroidal
magnetic field configuration in the Poynting flux, but adopting a
different magnetic configuration would not significantly affect the
estimate of $t_\Delta$.

After the forward shock crosses the ejecta, the forward shock
ploughs into the ambient medium. The dynamics of the blastwave
during this stage is defined by energy conservation\footnote{The
accurate expression for Eq.\ref{Dyn} should be
$L_{\rm{0}}t=(\gamma-\gamma_{\rm ej,0})M_{\rm
ej}c^2+(\gamma^{2}-1)M_{\rm sw}c^2$, where $\gamma_{\rm ej,0}$ is
the initial Lorenz factor for the ejecta, which we take as unity for
convenience. }
\begin{eqnarray}\label{Dyn}
L_{\rm{0}}t=(\gamma-1)M_{\rm ej}c^2+(\gamma^{2}-1)M_{\rm sw}c^2,
\end{eqnarray}
where $M_{\rm sw}=\frac{4\pi}{3}R^3nm_p$ is the swept mass from the
interstellar medium.  Initially, $(\gamma-1)M_{\rm
ej}c^2\gg(\gamma^{2}-1)M_{\rm sw}c^2$, so the kinetic energy of the
ejecta would increase linearly with time until $t={\rm min}(T_{\rm
sd}, T_{\rm dec})$, where the deceleration timescale $T_{\rm{dec}}$
is defined by the condition $(\gamma-1)M_{\rm
ej}c^2=(\gamma^{2}-1)M_{\rm sw}c^2$. By setting $T_{\rm{dec}}\sim
T_{\rm sd}$, we can derive a critical ejecta mass
\begin{eqnarray}
M_{\rm ej,c,1}\sim
10^{-3}M_{\odot}n^{1/8}I_{45}^{5/4}L_{0,47}^{-3/8}
P_{0,-3}^{-5/2}\xi^{5/4}\nonumber\\~~~~~~~~~~~~\sim 10^{-3}
M_{\odot}n^{1/8}I_{45}^{5/4}B^{-3/4}_{p,14}R_{6}^{-9/4}P_{0,-3}^{-1}
\xi^{7/8}
\end{eqnarray}
which separate regimes with different blastwave dynamics. For a
millisecond massive magnetar, the parameters $I_{45}$, $R_6$,
$P_{0,-3}$ are all essentially fixed values. The dependence on $n$
is very weak (1/8 power), so the key parameters that determine the
blastwave parameters are the ejecta mass $M_{\rm ej}$ and the
magnetar injection luminosity $L_0$ (or the magnetic field strength
$B_p$). If $M_{\rm ej} < M_{\rm ej,c,1}$ (or $T_{\rm dec} < T_{\rm
sd}$), the ejecta can be accelerated linearly until the deceleration
radius, after which the blastwave decelerates, but still with
continuous energy injection until $T_{\rm sd}$. Conversely, in the
opposite regime ($M_{\rm ej} > M_{\rm ej,c,1}$ or $T_{\rm sd} <
T_{\rm dec}$), the blastwave is only accelerated to $T_{\rm sd}$,
after which it coasts before decelerating at $T_{\rm dec}$. In the
intermediate regime of $M_{\rm ej} \sim M_{\rm ej,c,1}$ (or $T_{\rm
dec} \sim T_{\rm sd}$), the blastwave shows a decay after being
linearly accelerated.

There is another critical ejecta mass, which defines whether the
blastwave can reach a relativistic speed. This is defined by $E_{\rm
rot}\xi = 2 (\gamma-1) M_{\rm ej} c^2$. Defining a relativistic
ejecta as $\gamma-1 > 1$, this second critical ejecta mass is
\begin{equation}
 M_{\rm ej,c,2} \sim 6\times 10^{-3} M_\odot I_{45} P_{0,-3}^{-2} \xi.
\end{equation}
An ejecta heavier than this would not be accelerated to a
relativistic speed.

Below we discuss four dynamical regimes.

\begin{figure}[h!!!]
\begin{minipage}[b]{0.5\textwidth}
\centering \psfig{file=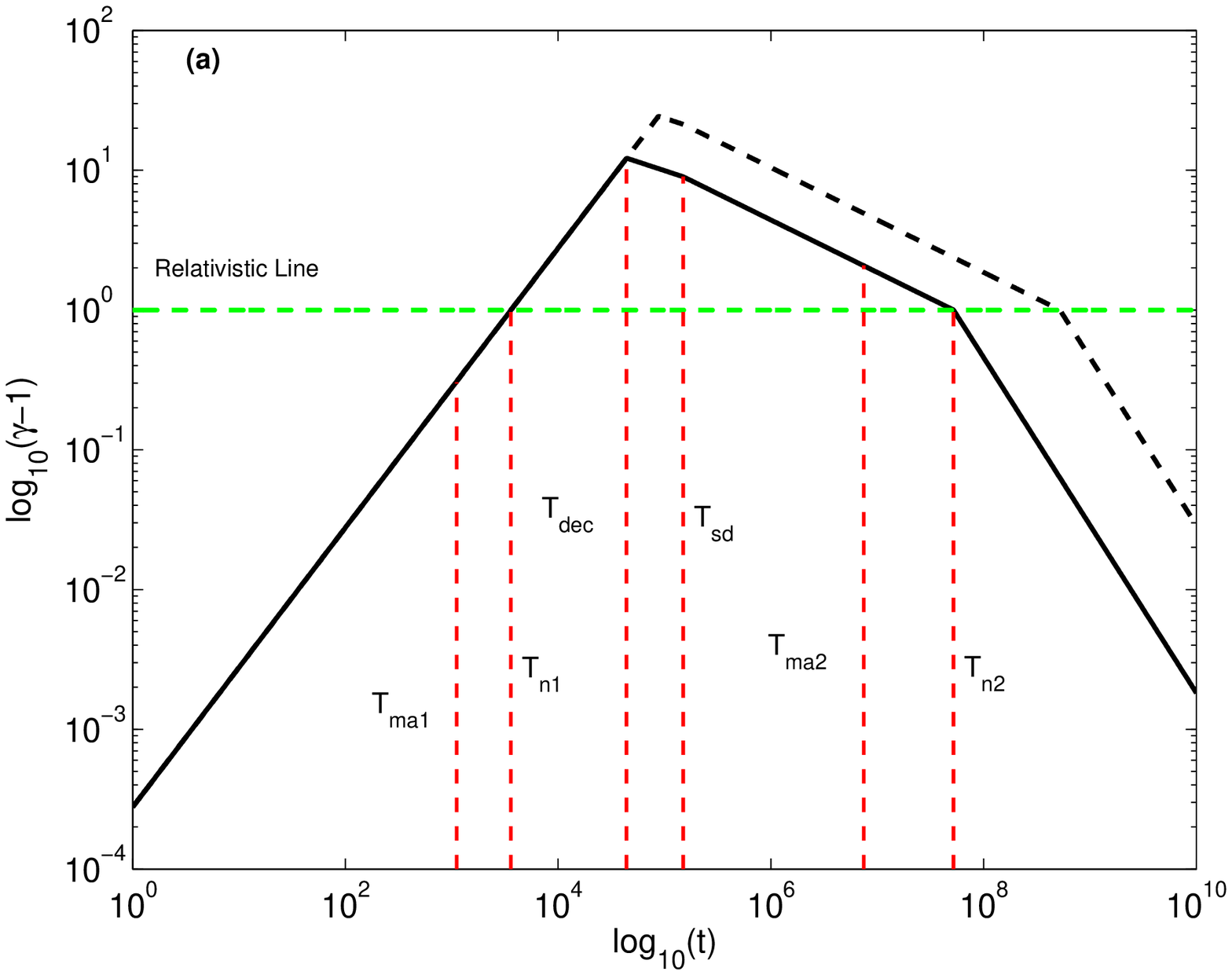,width=3in}
\end{minipage}%
\begin{minipage}[b]{0.5\textwidth}
\centering \psfig{file=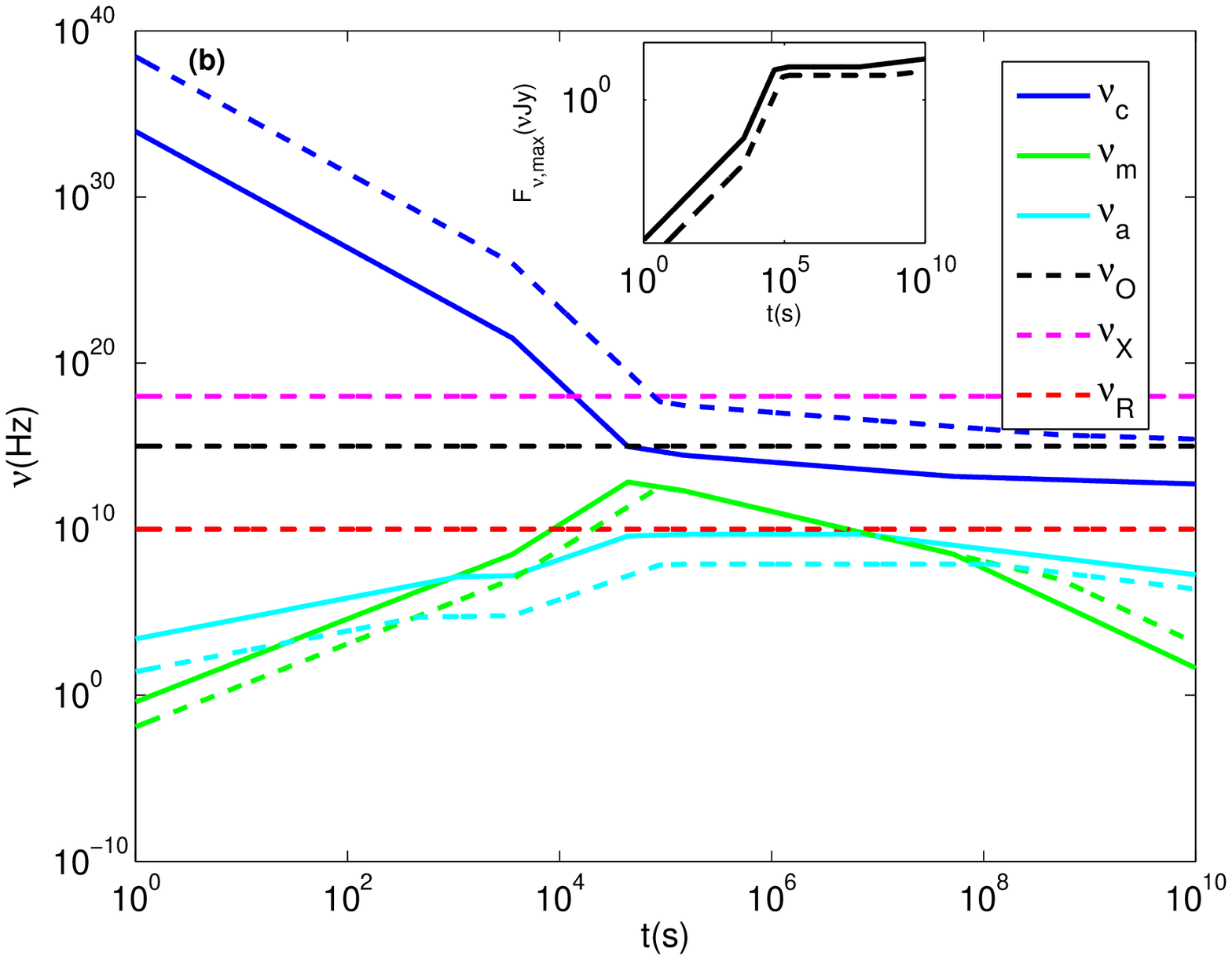,width=3in}
\end{minipage}\\
\begin{minipage}[b]{0.5\textwidth}
\centering \psfig{file=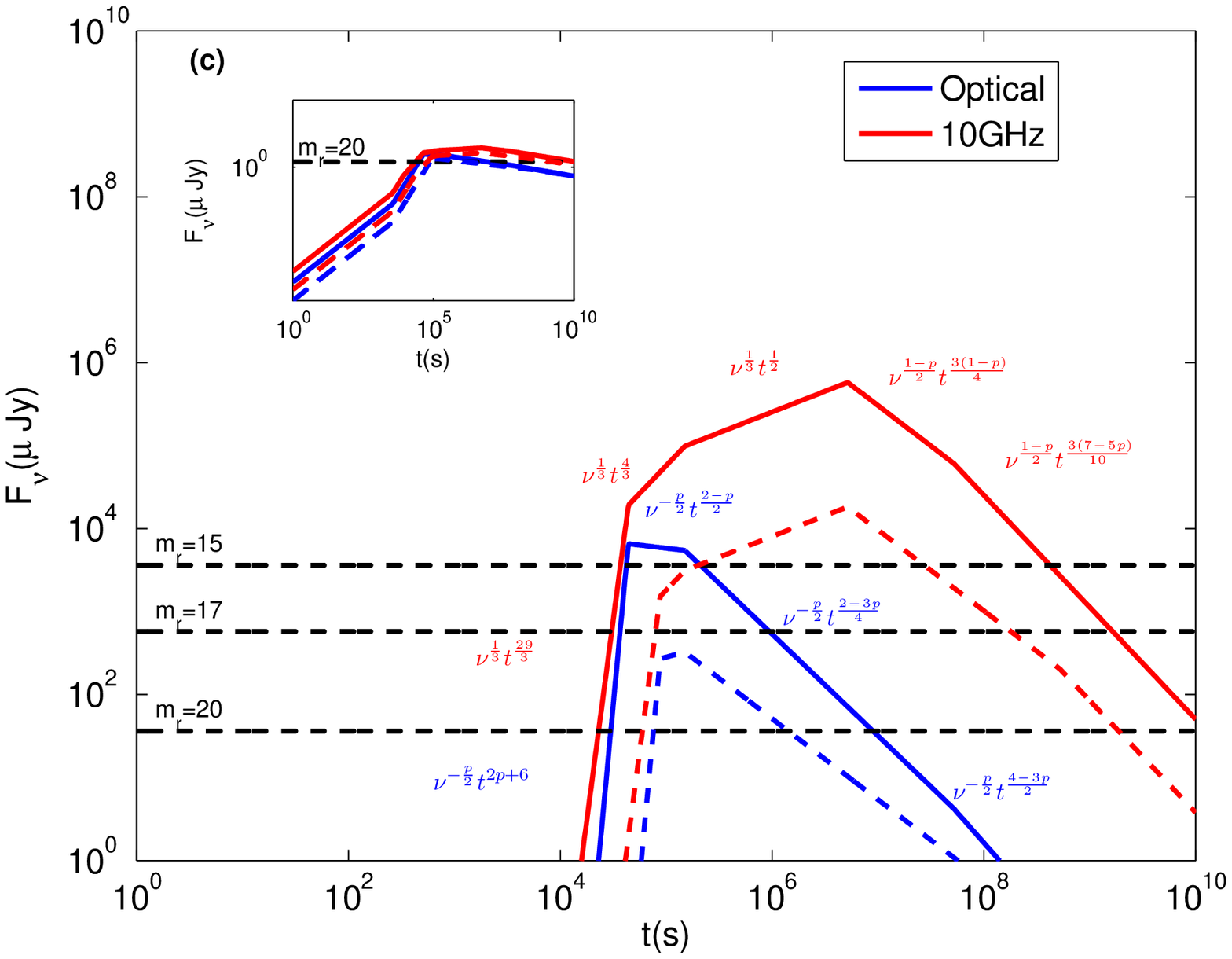,width=3in}
\end{minipage}%
\begin{minipage}[b]{0.5\textwidth}
\centering \psfig{file=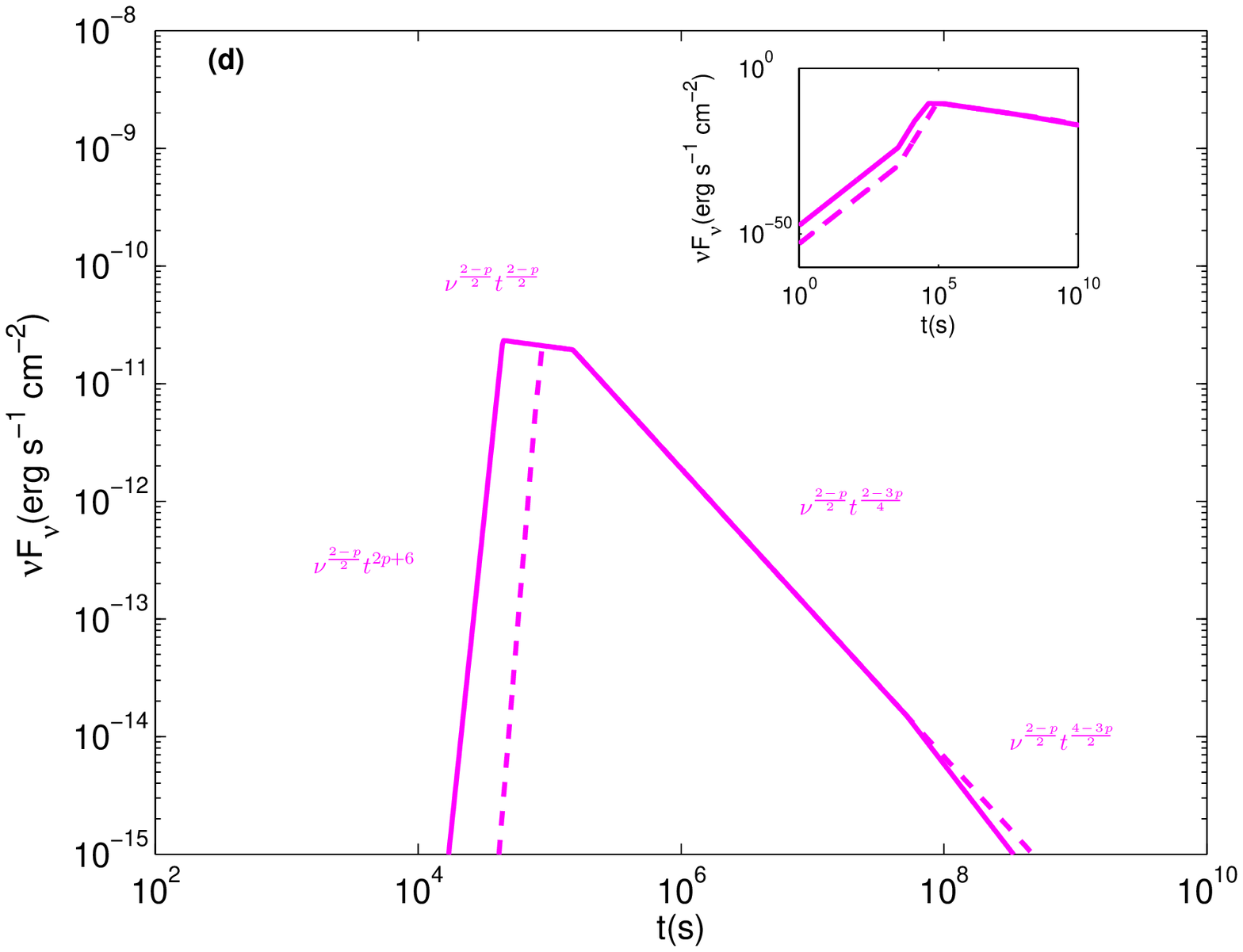,width=3in}
\end{minipage}\\
      \caption{Calculation results for Case I:
$L_{0}\sim10^{47}~{\rm erg ~s^{-1}}$ and $M_{\rm
ej}\sim10^{-4}M_{\odot}$(for all the examples, we adopt
$\xi=0.5,~p=2.3$).
       (a) The dynamical evolution of the parameter $(\gamma-1)$;
       (b) Temporal evolutions of the characteristic frequencies
$\nu_{\rm{a}}$, $\nu_{\rm{m}}$, and $\nu_{\rm{c}}$, and the peak
flux density $F_{\rm{\nu,max}}$;
       (c) Analytical light curve in R-band (blue) and 10 GHz radio band (red);
       (d) Analytical light curve in X-ray band.
       The solid and dashed lines are for $n=1$ cm$^{-3}$ and $n=10^{-3}$ cm$^{-3}$, respectively. In (c) and (d), we mark the spectral
and temporal indices for each segment of the light curves for $n=1$
cm$^{-3}$. The main figures denote the time regimes when the light
curves are detectable. The insets show the full light curves for
completeness. Both X-ray and optical light curves reach their peaks
around $10^{4}$ s, and remain detectable in years. The radio light
curve peaks around $10^{7}$ s, and lasts even longer. The peak flux
for X-ray, optical and radio could be as bright as $10^{-11}~{\rm
erg ~ s^{-1} cm^{-2}}$, $10 ~{\rm mJy}$ and $ {\rm Jy}$,
respectively.}
           \label{I}
            \end{figure}

\emph{Case I: $M_{\rm ej} < M_{\rm ej,c,1}$ or $T_{\rm sd}>T_{\rm
dec}$.} This requires both a small $L_0$ (or low $B_p$) and a small
$M_{\rm ej}$. We take an example with $L_0 \sim 10^{47}~{\rm
erg~s^{-1}}$ ($B_p \sim 10^{14}$ G) and $M_{\rm ej}\sim 10^{-4}
M_{\odot}$. To describe the dynamics in such a case, besides the
spin down timescale $T_{\rm{sd}}$, we need three more characteristic
time scales and the Lorentz factor value at the deceleration time
\begin{eqnarray}
&&T_{\rm{dec}}\sim4.4\times10^{4}~{\rm s}~L_{0,47}^{-7/10}M_{\rm ej,-4}^{4/5}n^{-1/10}\nonumber\\
&&T_{\rm{N1}}\sim3.6\times10^{3}~{\rm s}~L_{0,47}^{-1}M_{\rm ej,-4}^{}\nonumber\\
&&T_{\rm{N2}}\sim4.5\times10^{7}~{\rm s}~L_{0,47}^{1/3}n^{-1/3}T_{\rm sd,5}^{1/3}\nonumber\\
&&\gamma_{\rm{dec}}\sim12.2L_{0,47}^{3/10}M_{\rm
ej,-4}^{-1/5}n^{-1/10}+1
\end{eqnarray}
where $T_{\rm{N1}}, T_{\rm{N2}}$ are the two time scales when the
blastwave passes the non-relativistic to relativistic transition
line $\gamma-1=1$ during the acceleration and deceleration phases.
With these parameters, one can characterize the dynamical evolution
of the blastwave (Fig.\ref{I}a), as shown in Table 1.
Based on the dynamics, we can quantify the temporal evolution of
synchrotron radiation characteristic frequencies $\nu_{\rm{a}},
\nu_{\rm{m}}, \nu_{\rm{c}}$, and the peak flux, $F_{\rm{\nu,max}}$.
The evolutions of the characteristic frequencies are presented in
Fig.\ref{I}b and collected in Table 2.

Following the standard procedure in \cite{sari98}, we derive the
synchrotron radiation characteristic frequencies and the peak flux
density at $T_{\rm{dec}}$,
\begin{eqnarray}
&&\nu_{\rm{a,dec}}\sim5.0\times10^{8}~{\rm Hz}~L_{0,47}^{3/50}M_{\rm ej,-4}^{4/25}n^{29/50}\epsilon_{e,-1}^{-1}\epsilon_{B,-2}^{1/5}\nonumber\\
&&~~~~~~~~~~~~~~~~~~~~~~~\times\left(\frac{p-1}{p-2}\right)(p+1)^{3/5}f(p)^{3/5}\nonumber\\
&&\nu_{\rm{m,dec}}\sim1.3\times10^{14}~{\rm Hz}~L_{0,47}^{6/5}M_{\rm ej,-4}^{-4/5}n^{1/10}\epsilon_{e,-1}^{2}\epsilon_{B,-2}^{1/2}\left(\frac{p-2}{p-1}\right)^{2}\nonumber\\
&&\nu_{\rm{c,dec}}\sim9.6\times10^{14}~{\rm Hz}~L_{0,47}^{1/5}M_{\rm ej,-4}^{-4/5}n^{-9/10}\epsilon_{B,-2}^{-3/2}\nonumber\\
&&F_{\rm{\nu,max,dec}}\sim1.7\times10^{5}~\mu {\rm
Jy}~L_{0,47}^{3/10}M_{\rm
ej,-4}^{4/5}n^{2/5}\epsilon_{B,-2}^{1/2}D_{27}^{-2}
\end{eqnarray}
where
$f(p)=\frac{\Gamma(\frac{3p+22}{12})\Gamma(\frac{3p+2}{12})}{\Gamma(\frac{3p+19}{12})\Gamma(\frac{3p-1}{12})}$.
With the temporal evolution power law indices of these parameters
(Table 2), one can calculate the X-ray, optical and radio afterglow
lightcurves. Notice that there are two more temporal segments listed
in Table 2, since $\nu_a$ crosses $\nu_m$ twice at
\begin{eqnarray}
&&T_{\rm{ma1}}\sim1.4\times10^{2}~{\rm s}~L_{0,47}^{-5/4}M_{\rm ej,-4}^{5/4}n^{1/8}\epsilon_{e,-1}^{-5/4}\epsilon_{B,-2}^{-1/8}\left(\frac{p-2}{p-1}\right)^{-5/4}\nonumber\\
&&~~~~~~~~~~~~~~~~~~~~~~~\times(p+1)^{1/4}f(p)^{1/4},\nonumber\\
&&T_{\rm{ma2}}\sim1.9\times10^{8}~{\rm
s}~L_{0,47}^{1/5}n^{-2/5}T_{\rm
sd,5}^{1/5}\epsilon_{e,-1}^{2}\epsilon_{B,-2}^{1/5}\left(\frac{p-2}{p-1}\right)^{2}\nonumber\\
&&~~~~~~~~~~~~~~~~~~~~~~~\times(p+1)^{-2/5}f(p)^{-2/5},
\end{eqnarray}
respectively. We present the light curves in X-ray (Fig.\ref{I}d),
optical and radio (10GHz) band (Fig.\ref{I}c). The distance is taken
as 300 Mpc, the detection horizon of Advanced LIGO.

\emph{Case II: $M_{\rm ej} \sim M_{\rm ej,c,1}$ or $T_{\rm sd} \sim
T_{\rm dec}$.} The dynamics and the expressions of the
characteristic parameters become simpler:
\begin{eqnarray}
\label{dynII}
&&T_{\rm{dec}}\sim T_{\rm sd}\nonumber\\
&&T_{\rm{N1}}\sim 12~{\rm s}~\xi^{-1}M_{\rm ej,-4}^{}T_{\rm sd,3}\nonumber\\
&&T_{\rm{N2}}\sim 1.3\times10^{8}~{\rm s}~\xi^{8/3}M_{\rm ej,-4}^{-8/3}T_{\rm sd,3}\nonumber\\
&&\gamma_{\rm{sd}}\sim 83.3\xi M_{\rm ej,-4}^{-1}+1
\end{eqnarray}
The temporal indices of the evolutions of $\nu_{\rm{a}},
\nu_{\rm{m}}, \nu_{\rm{c}}, F_{\rm{\nu,max}}$ are listed in Table 2,
and the expressions of $\gamma$ and $R$ are shown in Table 1.

As examples, we consider $L_0 \sim 10^{49}~{\rm erg~s^{-1}}$ ($B_{p}
\sim 10^{15}$ G) vs. $M_{\rm ej}\sim 10^{-4} M_{\odot}$, which
satisfies $T_{\rm{sd}} \sim T_{\rm{dec}}$.

Similarly to Case I, we have
\begin{eqnarray}
\label{dynIIamc1}
&&\nu_{\rm{a,sd}}\sim2.2\times10^{9}~{\rm Hz}~\xi^{11/5}L_{0,49}^{-3/5}M_{\rm ej,-4}^{-8/5}n^{4/5}\epsilon_{e,-1}^{-1}\epsilon_{B,-2}^{1/5}\nonumber\\
&&~~~~~~~~~~~~~~~~~~~~~\times\left(\frac{p-1}{p-2}\right)(p+1)^{3/5}f(p)^{3/5}\nonumber\\
&&\nu_{\rm{m,sd}}\sim2.7\times10^{17}~{\rm Hz}~\xi^{4}M_{\rm ej,-4}^{-4}n^{1/2}\epsilon_{e,-1}^{2}\epsilon_{B,-2}^{1/2}\left(\frac{p-2}{p-1}\right)^{2}\nonumber\\
&&\nu_{\rm{c,sd}}\sim8.6\times10^{14}~{\rm Hz}~\xi^{-4}M_{\rm
ej,-4}^{4}n^{-3/2}T_{\rm
sd,3}^{-2}\epsilon_{B,-2}^{-3/2}\nonumber\\
&&F_{\rm{\nu,max,sd}}\sim2.4\times10^{8}~\mu {\rm
Jy}~\xi^{11}L_{0,49}^{-3}M_{\rm ej,-4}^{-8}n^{3/2}\epsilon_{B,-2}^{1/2}D_{27}^{-2}\nonumber\\
&&T_{\rm{ma1}}\sim1.4\times10^{-1}~{\rm
s}~\xi^{-1}L_{0,49}^{-1/4}M_{\rm ej,-4}^{5/4}n^{1/8}T_{\rm
sd,3}\epsilon_{e,-1}^{-5/4}\epsilon_{B,-2}^{-1/8}\nonumber\\
&&~~~~~~~~~~~~~~~~~~~~~\times\left(\frac{p-2}{p-1}\right)^{-5/4}(p+1)^{1/4}f(p)^{1/4}\nonumber\\
&&T_{\rm{ma2}}\sim2.5\times10^{8}~{\rm
s}~\xi^{6/5}L_{0,49}^{2/5}M_{\rm ej,-4}^{-8/5}n^{-1/5}T_{\rm
sd,3}^{}\epsilon_{e,-1}^{2}\epsilon_{B,-2}^{1/5}\nonumber\\
&&~~~~~~~~~~~~~~~~~~~~~\times\left(\frac{p-2}{p-1}\right)^{2}(p+1)^{-2/5}f(p)^{-2/5}
\end{eqnarray}

\begin{figure}[h!!!]
\begin{minipage}[b]{0.5\textwidth}
\centering \psfig{file=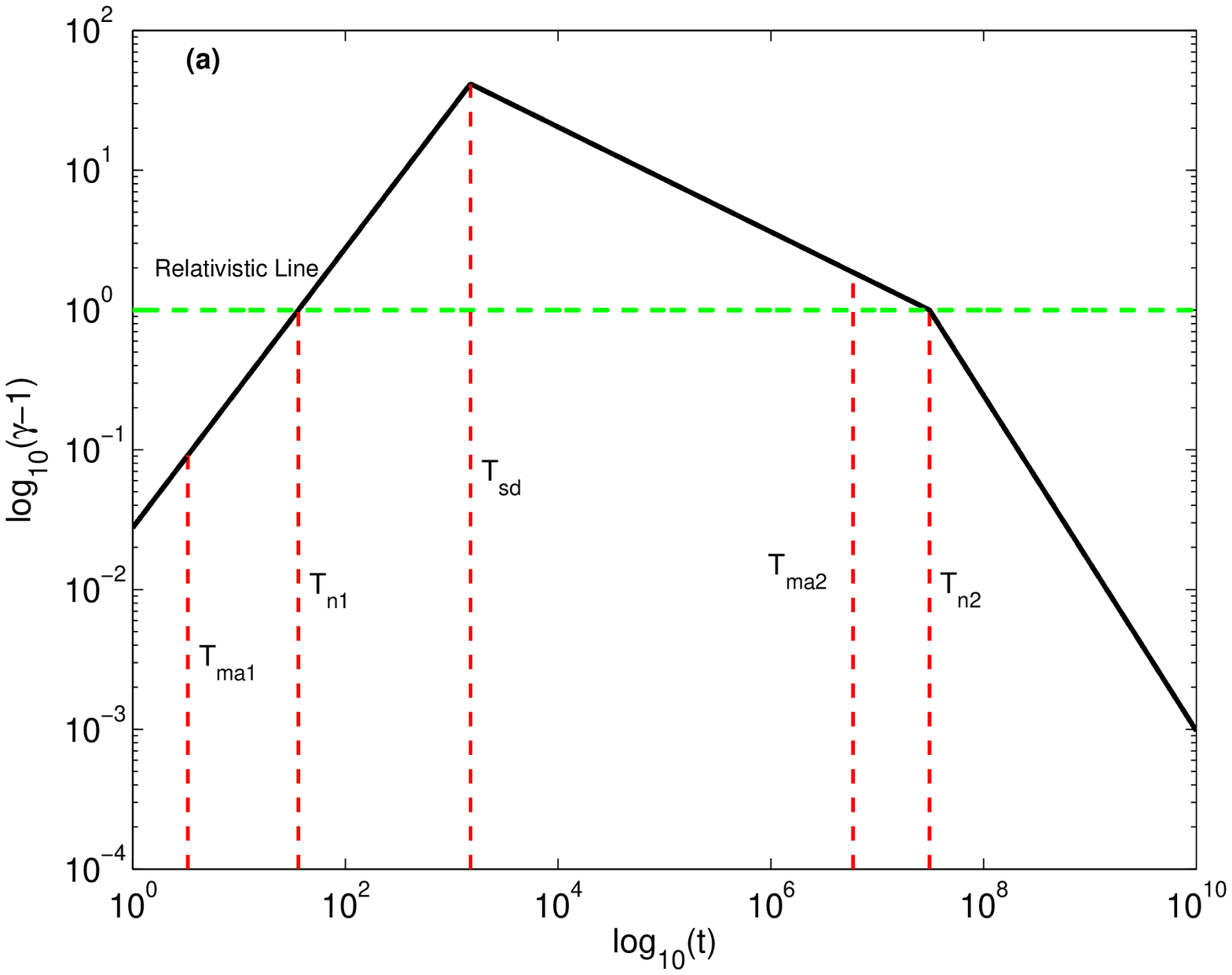,width=3in}
\end{minipage}%
\begin{minipage}[b]{0.5\textwidth}
\centering \psfig{file=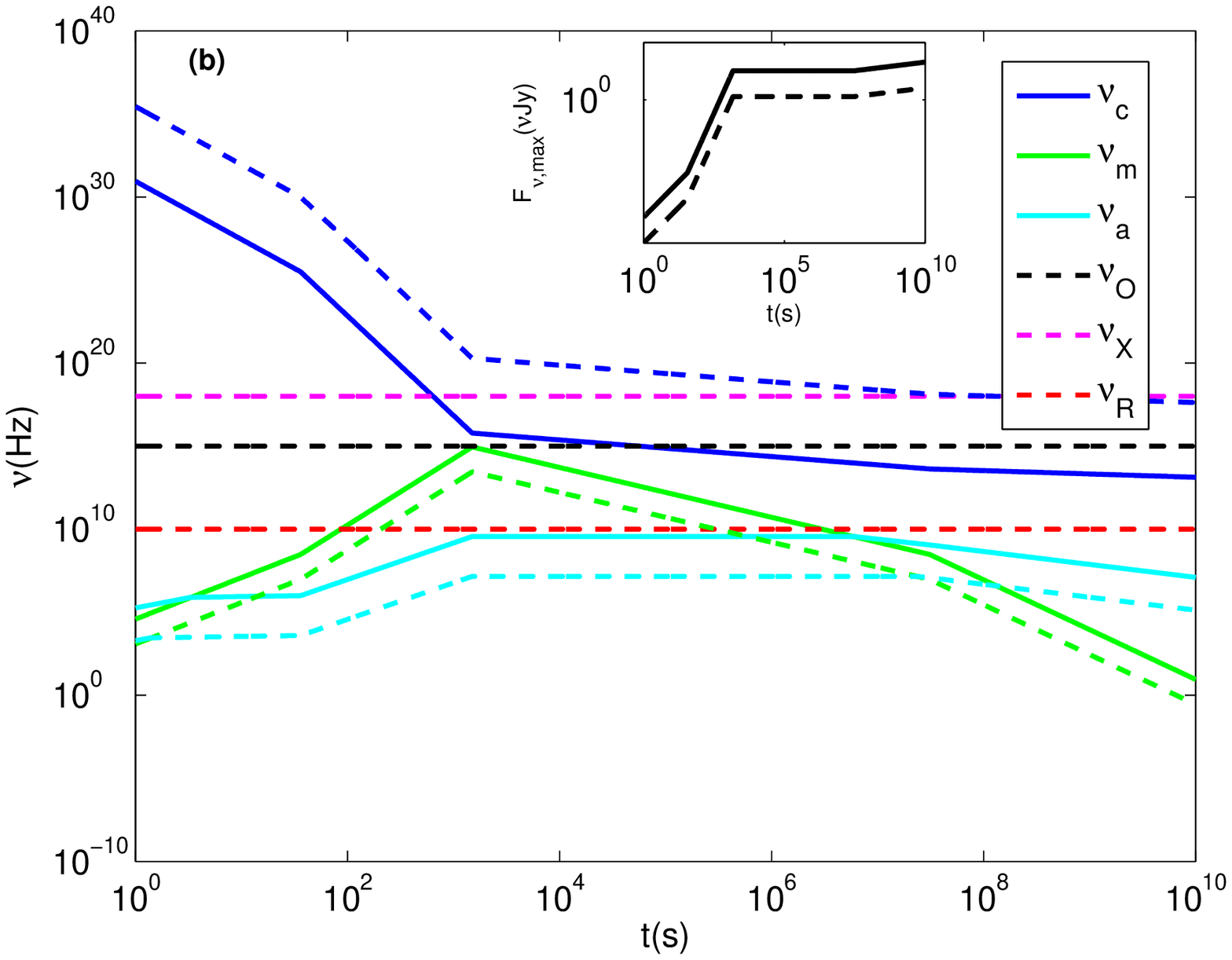,width=3in}
\end{minipage}\\
\begin{minipage}[b]{0.5\textwidth}
\centering \psfig{file=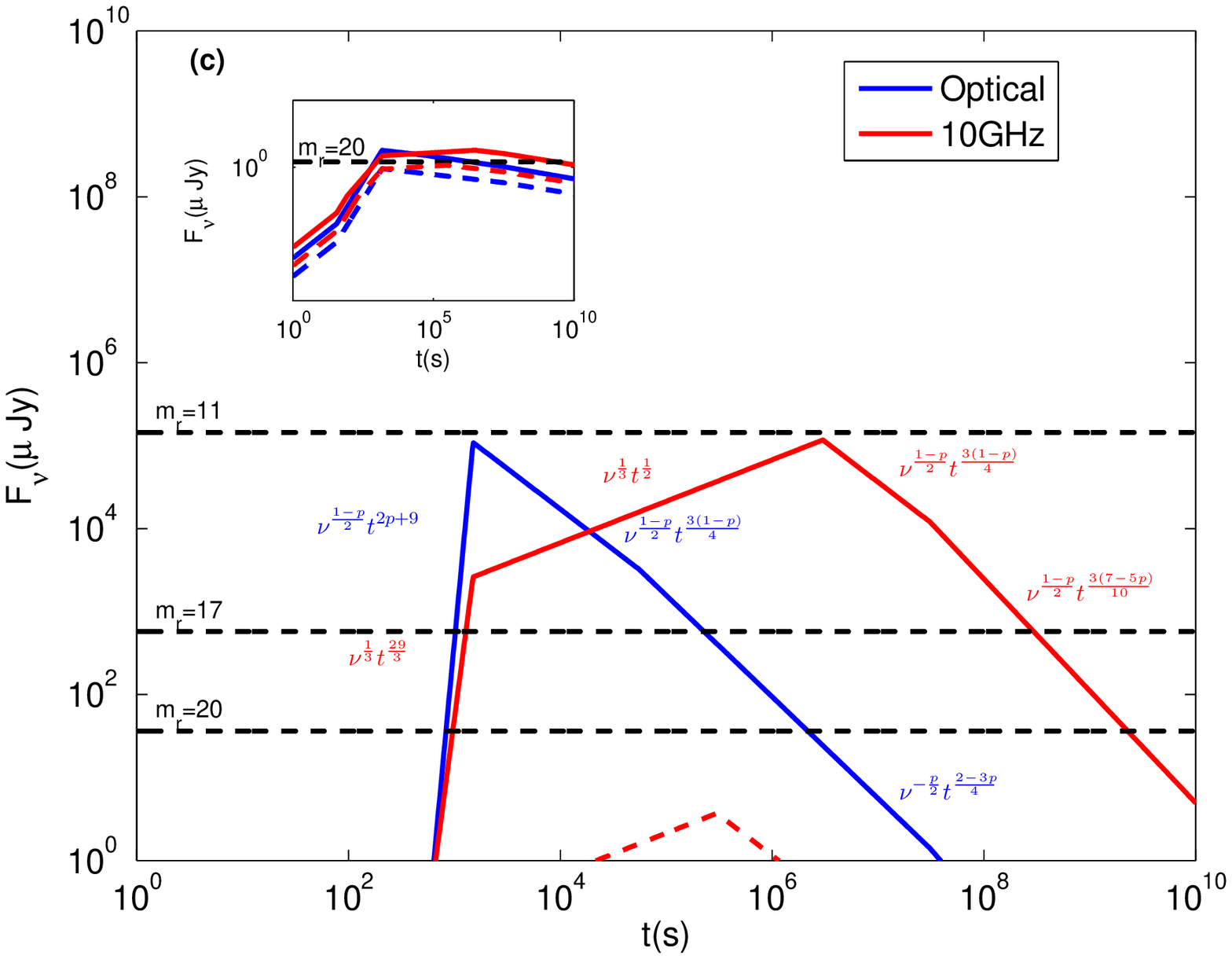,width=3in}
\end{minipage}%
\begin{minipage}[b]{0.5\textwidth}
\centering \psfig{file=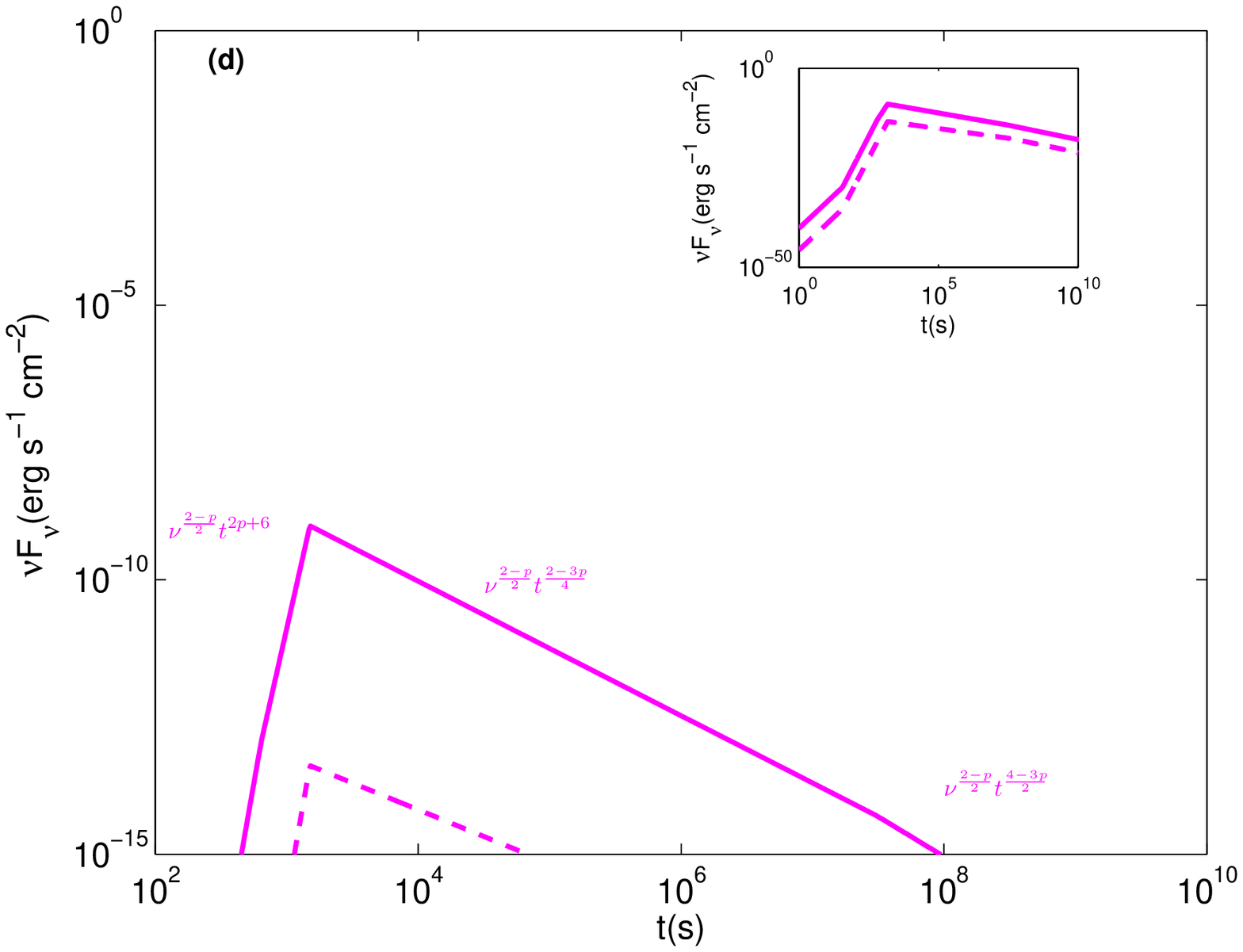,width=3in}
\end{minipage}\\
       \caption{Calculation results for Case II: $L_{0}\sim10^{49}~{\rm erg ~s^{-1}}$
and $M_{\rm ej}\sim10^{-4}M_{\odot}$. Captions are the same with
Figure 2. For $n=1$ cm$^{-3}$, both X-ray and optical light curve
reach their peaks around $10^{3}$ s, and the radio light curve peaks
around $10^{7}$ s. The peak flux of X-ray, optical, and radio is
$10^{-9} ~{\rm erg ~s^{-1} cm^{-2}}$, $100~{\rm mJy}$ and $100~{\rm
mJy}$, respectively. Taking R-band magnitude 20 and $10^{-15}~{\rm
erg ~s^{-1} cm^{-2}}$ as the detection limit, the durations of the
detectable optical and X-ray afterglow are $\sim 10^6$ s and $\sim
10^8$ s respectively. The radio afterglow lasts even longer. For
$n=10^{-3}$ cm$^{-3}$, the signals for X-ray is still detectable,
but with shorter durations, $\sim 10^5$ s.}
           \label{II}
            \end{figure}

The expressions of $\gamma$ and $R$ as well as the power-law indices for this
case are also presented in Table 1 and Table 2, respectively. The dynamics
typical frequency evolution, and the light curves are presented
in Fig.\ref{II}. We note that in this case (and case
III), the synchrotron radiation properties are very sensitive to
$M_{\rm ej}$ and $\xi$.

\emph{Case III: $M_{\rm ej,c,1} < M_{\rm ej} < M_{\rm ej,c,2}$}
($T_{\rm sd} < T_{\rm dec}$).  As an example, we take $B_{p} \sim
10^{15}$ G, and $M_{\rm ej}\sim 10^{-3} M_{\odot}$.

For this example, the dynamics and the expressions of the
characteristic parameters become
\begin{eqnarray}
\label{dynII}
&&T_{\rm{dec}}\sim 1.5\times 10^{4}~{\rm s}~\xi^{-7/3}M_{\rm ej,-3}^{8/3}n^{-1/3}\nonumber\\
&&T_{\rm{N1}}\sim 59.9~{\rm s}~\xi^{-1}M_{\rm ej,-3}^{}T_{\rm sd,3}\nonumber\\
&&T_{\rm{N2}}\sim 2.7\times 10^{7}~{\rm s}~\xi^{1/3}n^{-1/3}\nonumber\\
&&\gamma_{\rm{sd}}\sim 16.7\xi M_{\rm ej,-3}^{-1}+1
\end{eqnarray}
and
\begin{eqnarray}
\label{dynIIamc1}
&&\nu_{\rm{a,sd}}\sim1.6\times10^{8}~{\rm Hz}~\xi^{8/5}M_{\rm ej,-3}^{-8/5}n^{4/5}T_{\rm sd,3}^{3/5}\epsilon_{e,-1}^{-1}\epsilon_{B,-2}^{1/5}\left(\frac{p-1}{p-2}\right)\nonumber\\
&&~~~~~~~~~~~~~~~~~~~~~~\times(p+1)^{3/5}f(p)^{3/5}\nonumber\\
&&\nu_{\rm{m,sd}}\sim4.5\times10^{14}~{\rm Hz}~\xi^{4}M_{\rm ej,-3}^{-4}n^{1/2}\epsilon_{e,-1}^{2}\epsilon_{B,-2}^{1/2}\left(\frac{p-2}{p-1}\right)^{2}\nonumber\\
&&\nu_{\rm{c,sd}}\sim5.3\times10^{17}~{\rm Hz}~\xi^{-4}M_{\rm
ej,-3}^{4}n^{-3/2}T_{\rm
sd,3}^{-2}\epsilon_{B,-2}^{-3/2}\nonumber\\
&&F_{\rm{\nu,max,sd}}\sim6.5\times10^{2}~\mu {\rm
Jy}~\xi^{8}M_{\rm ej,-3}^{-8}n^{3/2}T_{\rm sd,3}^{3}\epsilon_{B,-2}^{1/2}D_{27}^{-2}\nonumber\\
&&T_{\rm{ma1}}\sim1.0~{\rm s}~\xi^{-5/4}M_{\rm
ej,-3}^{5/4}n^{1/8}T_{\rm
sd,3}^{5/4}\epsilon_{e,-1}^{-5/4}\epsilon_{B,-2}^{-1/8}\left(\frac{p-2}{p-1}\right)^{-5/4}\nonumber\\
&&~~~~~~~~~~~~~~~~~~~~~~\times(p+1)^{1/4}f(p)^{1/4}\nonumber\\
&&T_{\rm{ma2}}\sim9.9\times10^{7}~{\rm
s}~\xi^{1/5}n^{-2/5}\epsilon_{e,-1}^{2}\epsilon_{B,-2}^{1/5}\left(\frac{p-2}{p-1}\right)^{2}\nonumber\\
&&~~~~~~~~~~~~~~~~~~~~~~\times(p+1)^{-2/5}f(p)^{-2/5}
\end{eqnarray}

\begin{figure}[h!!!]
\begin{minipage}[b]{0.5\textwidth}
\centering \psfig{file=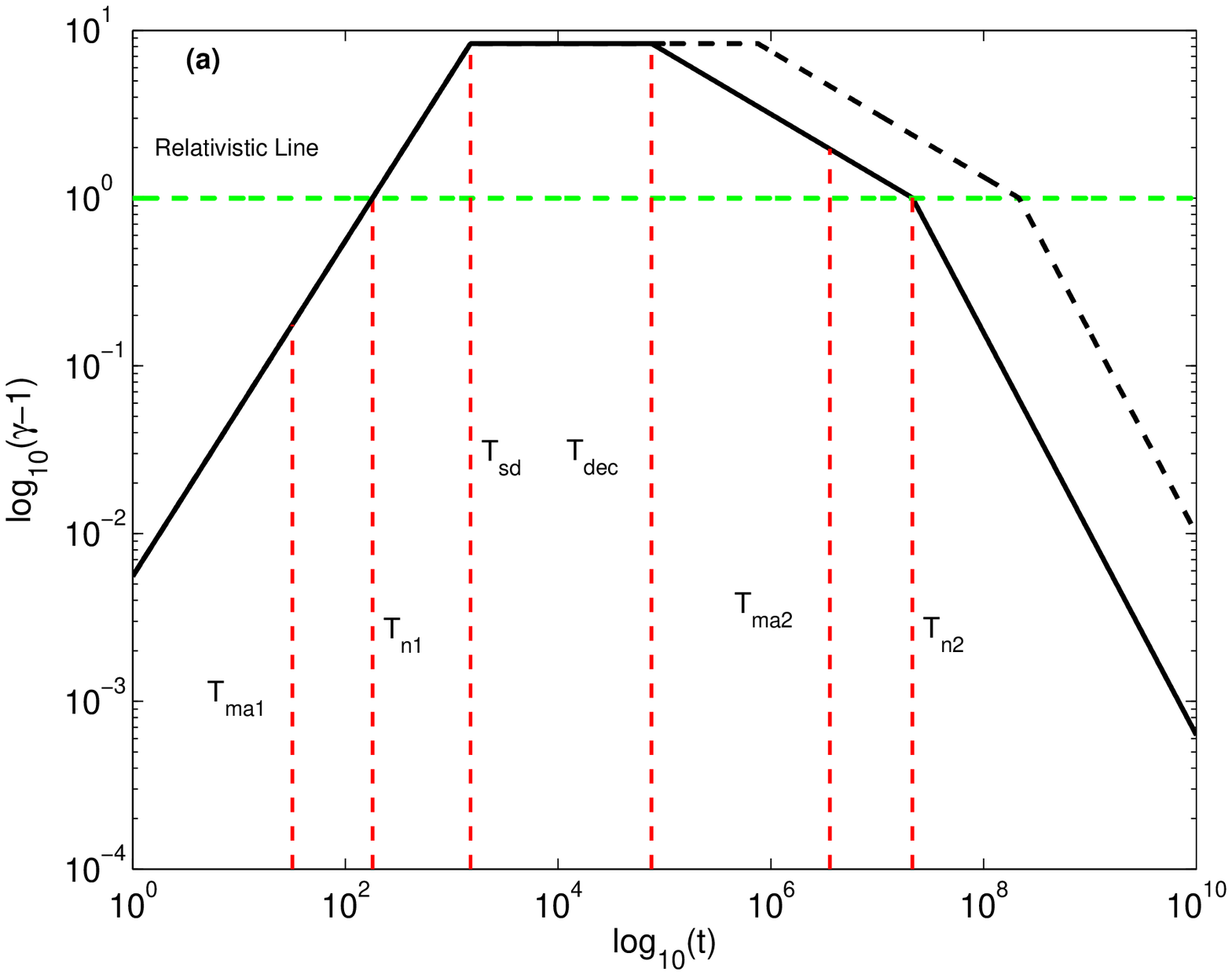,width=3in}
\end{minipage}%
\begin{minipage}[b]{0.5\textwidth}
\centering \psfig{file=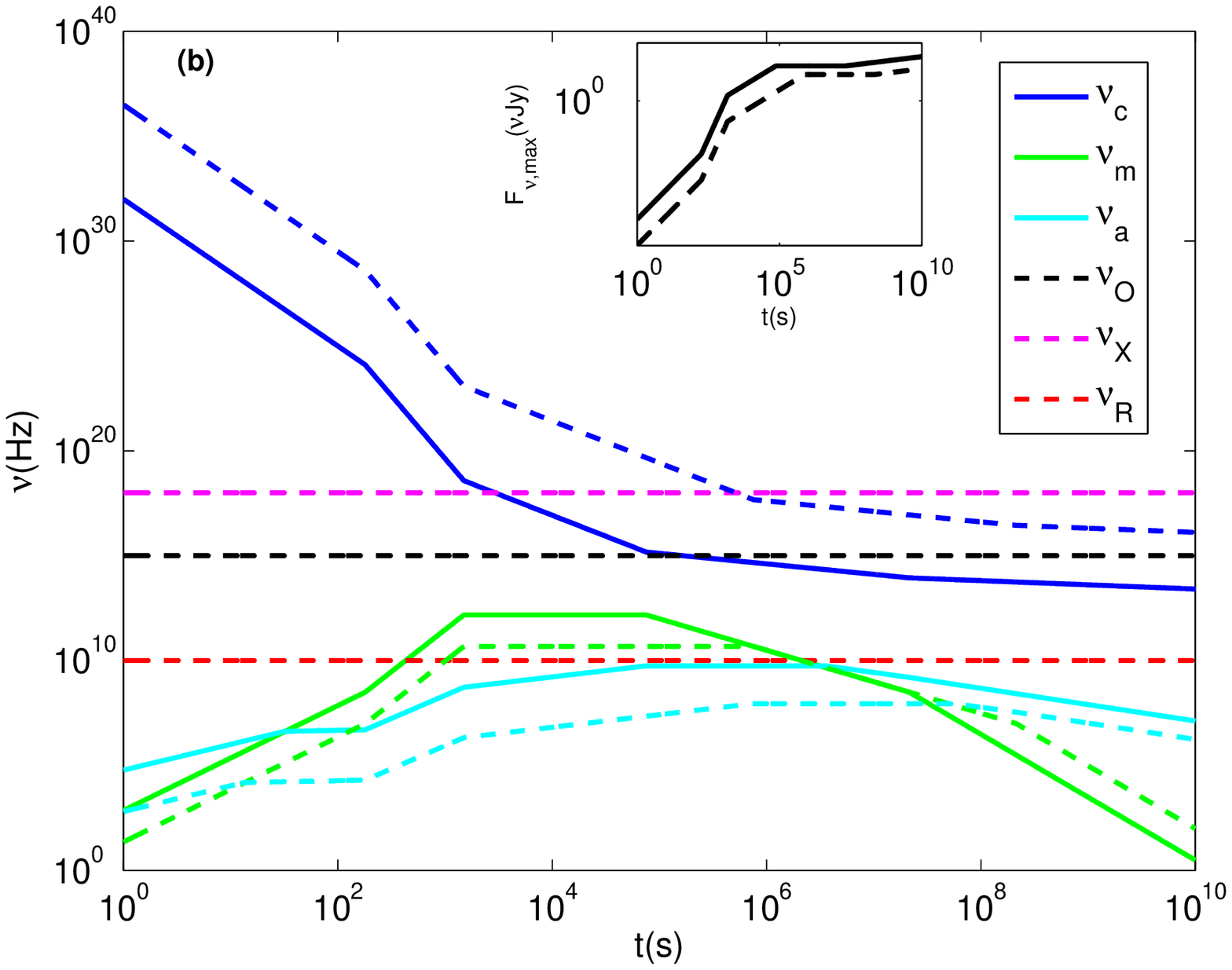,width=3in}
\end{minipage}\\
\begin{minipage}[b]{0.5\textwidth}
\centering \psfig{file=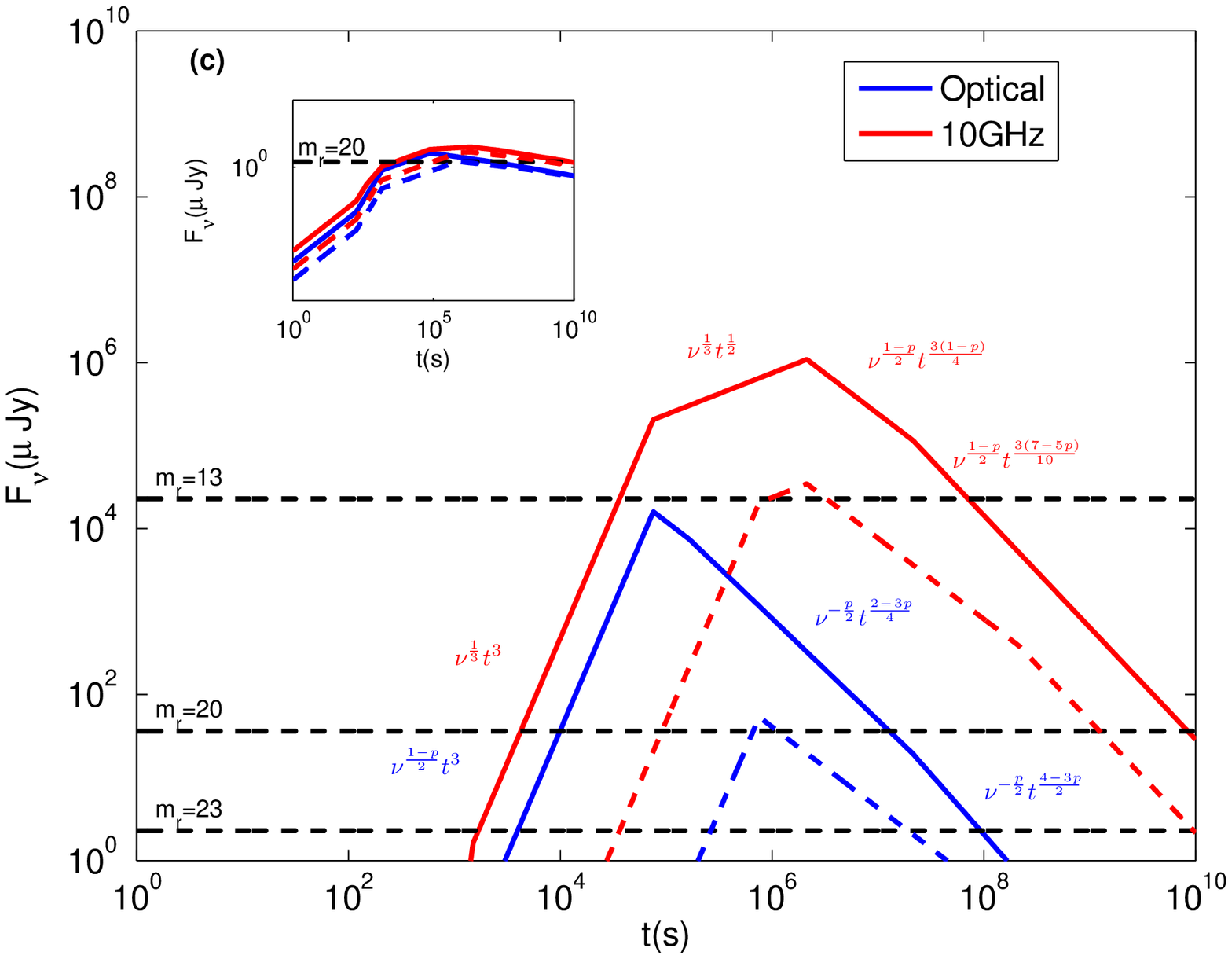,width=3in}
\end{minipage}%
\begin{minipage}[b]{0.5\textwidth}
\centering \psfig{file=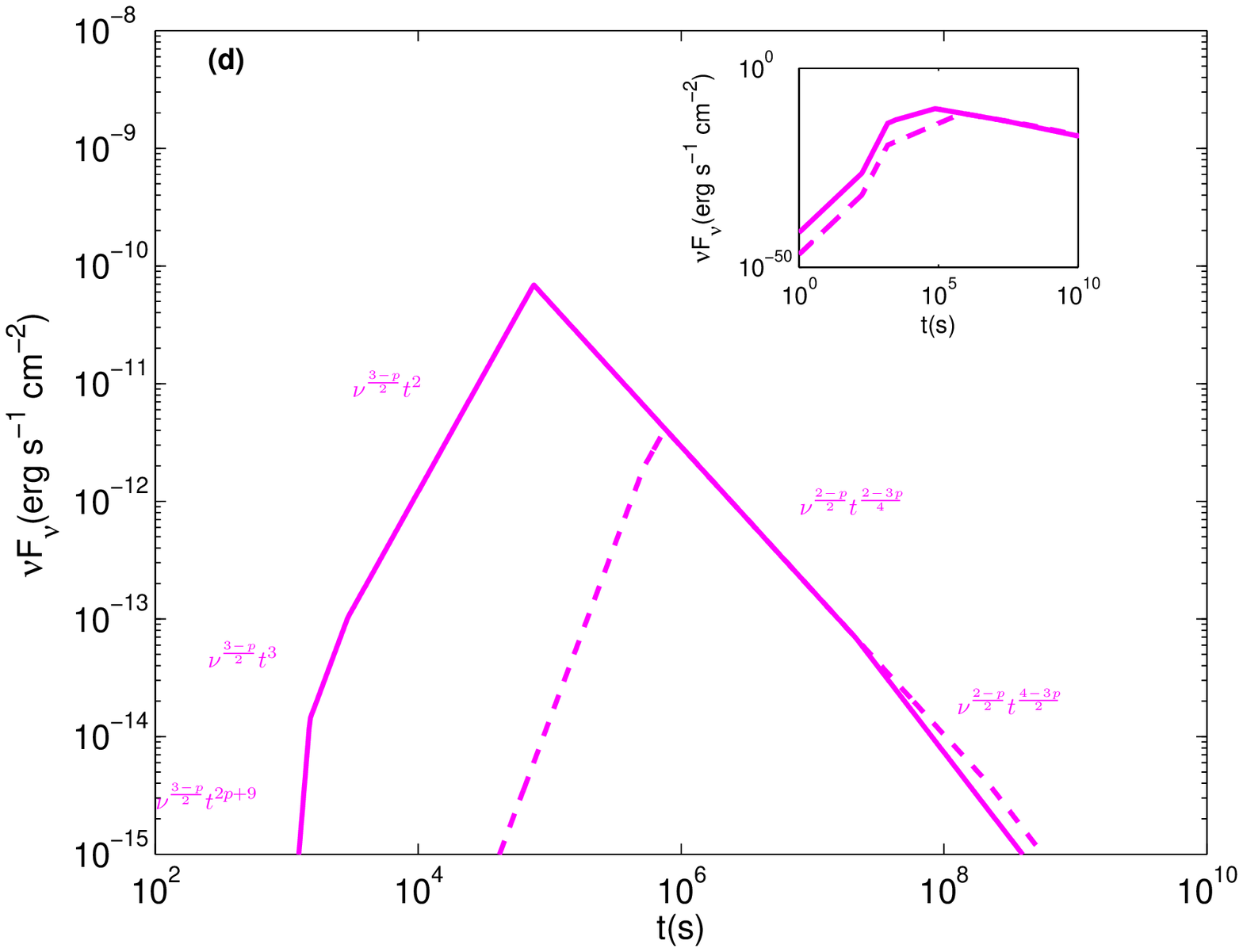,width=3in}
\end{minipage}\\
       \caption{Calculation results for Case III: $L_{0}\sim10^{49} ~{\rm erg ~s^{-1}}$
and $M_{\rm ej}\sim10^{-3}M_{\odot}$. Captions are the same with
Figure 2. For $n=1$ cm$^{-3}$, both X-ray and optical light curve
reach their peaks around $10^{5}$ s, and the radio light curve peaks
around $10^{6}$ s. The peak flux for X-ray, optical and radio is
$10^{-10} ~{\rm erg ~s^{-1} cm^{-2}}$, $10~{\rm mJy}$ and ${\rm
Jy}$, respectively. Taking R-band magnitude 20 and $10^{-15}~{\rm
erg s^{-1} cm^{-2}}$ as the detection limit, the durations of the
observable optical and X-ray afterglows are $\sim 10^{7}$ s and
$\sim 10^{8}$ s respectively. The radio duration lasts even longer.
If $n=10^{-3}$ cm$^{-3}$, the optical signal just reaches R-band
magnitude 20 around $\sim 10^{6}$ s, while the X-ray afterglow is
detectable with a duration $\sim 10^{8}$ s.}
           \label{III}
            \end{figure}

The power-law indices of various parameters for this case are also
collected in Table 2, and the dynamics, frequency evolutions, and
light curves are presented in Fig. \ref{III}.

\emph{Case IV: $M_{\rm ej} > M_{\rm ej,c,2}$.} In this case, the
blast wave never reaches a relativistic speed. The dynamics is
similar to Case III, with the coasting regime in the
non-relativistic phase. The dynamics for a non-relativistic ejecta
and its radio afterglow emission have been discussed in
\cite{nakar11}. Our Case IV resembles what is discussed in
\cite{nakar11}, but the afterglow flux is much enhanced because of a
larger total energy involved.

\section{Detectability and implications}

For all the cases, bright broadband EM afterglow emission signals
are predicted. The light curves typically show a sharp rise around
$T_{\rm{sd}}$, which coincides the ending time of the X-ray
afterglow signal discussed by \cite{zhang13} due to internal
dissipation of the magnetar wind. The X-ray afterglow luminosity
predicted in our model is generally lower than that of the internal
dissipation signal, but the optical and radio signals are much
brighter. In some cases, the R-band magnitude can reach 11th at the
300 Mpc, if $M_{\rm ej}$ is small enough (so that the blastwave has
a high Lorentz factor) and the medium density is not too low. The
duration of detectable optical emission ranges from $10^3$ seconds
to year time scale. The radio afterglow can reach the Jy level for
an extended period of time, with peak reached in the year time
scale. These signals can be readily picked up by all-sky optical
monitors, and radio surveys. The X-ray afterglow can be also picked
up by large field-of-view imaging telescopes such as ISS-Lobster.

Since these signals are originated from interaction between the
magnetar wind and the ejecta in the equatorial directions, they are
not supposed to be accompanied with short GRBs, and some
internal-dissipation X-ray afterglows \citep{zhang13} in the free
wind zone. Due to a larger solid angle, the event rate for this
geometry (orange observer in Fig.1) should be higher than the other
two geometries (red and yellow observers in Fig.1). However, the
brightness of the afterglow critically depends on the unknown
parameters such as $M_{\rm ej}$, $B_p$ (and hence $L_0$), and $n$.
The event rate also crucially depends on the event rate of NS-NS
mergers and the fraction of mergers that leave behind a massive
magnetar rather than a black hole.

This afterglow signal is much stronger than the afterglow signal due
to ejecta-medium interaction with a black hole as the post-merger
product \citep{nakar11}. The main reason is the much larger energy
budget involved in the magnetar case. Since the relativistic phase
can be achieved, both X-ray and optical afterglows are detectable,
which peak around the magnetar spindown time scale ($10^3 - 10^5$
s). The radio peak is later similar to the black hole case
\citep{nakar11}, but the radio afterglow flux is also much brighter
(reaching Jy level) due to a much larger energy budget involved. The
current event rate limit of $>350$ mJy radio transients in the
minutes-to-days time scale at 1.4 GHz is $<6\times 10^{-4} ~{\rm
degree}^{-2}~{\rm yr}^{-1}$ \citep{bower11}, or $< 20~{\rm yr}$ all
sky. In view of the large uncertainties in the NS-NS merger rate and
the fraction of millisecond magnetar as the post-merger product, our
prediction is entired consistent with this upper limit. Because of
their brightness, these radio transients can be detected outside the
Advanced LIGO horizon, which may account for some sub-mJy radio
transients discovered by VLA \citep{bower07}.

Recently, \cite{kyutoku12} proposed another possible EM counterpart
of GWB with a wide solid angle. They did not invoke a long-lasting
millisecond magnetar as the merger product, but speculated that
during the merger process, a breakout shock from the merging neutron
matter would accelerate a small fraction of surface material, which
reaches a relativistic speed. Such an outflow would also emit
broad-band synchrotron emission by shocking the surrounding medium.
Within that scenario, the predicted peak flux is lower and the
duration is shorter than the electro-magnetic signals predicted in
\citep{zhang13} and this work, due to a much lower energy carried by
the outflow.

Detecting the GWB-associated bright signals as discussed in this
paper would unambiguously confirm the astrophysical origin of
GWBs. Equally importantly, it would suggest that NS-NS mergers leave
behind a hyper-massive neutron star, which gives an important
constraint on the neutron star equation of state. With the GWB data,
one can infer the information of the two NSs involved in the merger.
Modeling afterglow emission can give useful constraints on the
ejected mass $M_{\rm ej}$ and the properties of the postmerger
compact objects. Therefore, a combination of GWB and afterglow
information would shed light into the detailed merger physics, and
in particular, provide a probe of massive millisecond magnetars and
stiff equations of state for neutron matter.

\acknowledgments We thank stimulative discussion with Yi-Zhong Fan
and Jian-Yan Wei. We acknowledge the National Basic Research Program
(``973" Program) of China under Grant No. 2009CB824800 and
2013CB834900. This work is also supported by the National Natural
Science Foundation of China (grant No. 11033002 \& 10921063) and by
NSF AST-0908362. XFW acknowledges support by the One-Hundred-Talents
Program and the Youth Innovation Promotion Association of Chinese
Academy of Sciences.

\clearpage

\begin{table}
\centering \caption{Expression of the Lorentz factor and radius as a
function of model parameters in different temporal regimes for all
dynamical cases.}
\begin{tabular}{llllllll}
\hline\hline

          &                                      &~~~~~~~~~~~~~~$\gamma$                                                         & ~~~~~~~~~~~~~~~ $R$                                                                   \\

\hline
          &  $t<T_{\rm{N1}}$                     &  $0.28L_{0,47}M_{\rm ej,-4}^{-1}t_{3}+1$                                       &  $3.2\times10^{13}L_{0,47}^{1/2}M_{\rm ej,-4}^{-1/2}t_{3}^{3/2}$                   \\
          &  $T_{\rm{N1}}<t<T_{\rm{dec}}$        &  $2.8L_{0,47}M_{\rm ej,-4}^{-1}t_{4}+1$                                        &  $4.6\times10^{15}L_{0,47}^{2}M_{\rm ej,-4}^{-2}t_{4}^{3}$                         \\
 Case I   &  $T_{\rm{dec}}<t<T_{\rm{sd}}$        &  $9.9L_{0,47}^{1/8}n^{-1/8}t_{5}^{-1/4}+1$                                     &  $5.9\times10^{17}L_{0,47}^{1/4}n^{-1/4}t_{5}^{1/2}$                               \\
          &  $T_{\rm{sd}}<t<T_{\rm N2}$          &  $4.2L_{0,47}^{1/8}T_{\rm sd,5}^{1/8}n^{-1/8}t_{6}^{-3/8}+1$                   &  $1.1\times10^{18}L_{0,47}^{1/4}T_{\rm sd,5}^{1/4}n^{-1/4}t_{6}^{1/4}$             \\
          &  $t>T_{\rm{N2}}$                     &  $0.4L_{0,47}^{2/5}T_{\rm sd,5}^{2/5}n^{-2/5}t_{8}^{-6/5}+1$                   &  $3.7\times10^{18}L_{0,47}^{1/5}T_{\rm sd,5}^{1/5}n^{-1/5}t_{8}^{2/5}$             \\
\hline
          &  $t<T_{\rm{N1}}$                     &  $0.08\xi M_{\rm ej,-4}^{-1}T_{\rm sd,3}^{-1}t_{}+1$                           &  $1.7\times10^{10}\xi^{1/2} M_{\rm ej,-4}^{-1/2}T_{\rm sd,3}^{-1/2}t_{}^{3/2}$     \\
 Case II  &  $T_{\rm{N1}}<t<T_{\rm{sd}}$         &  $83.3\xi M_{\rm ej,-4}^{-1}T_{\rm sd,3}^{-1}t_{3}+1$                          &  $4.2\times10^{17}\xi^{2} M_{\rm ej,-4}^{-2}T_{\rm sd,3}^{-2}t_{3}^{3}$            \\
          &  $T_{\rm{sd}}<t<T_{\rm N2}$          &  $14.8\xi M_{\rm ej,-4}^{-1}T_{\rm sd,3}^{3/8}t_{5}^{-3/8}+1$                  &  $1.3\times10^{18}\xi^{2} M_{\rm ej,-4}^{-2}T_{\rm sd,3}^{3/4}t_{5}^{1/4}$         \\
          &  $t>T_{\rm{N2}}$                     &  $1.4\xi^{16/5} M_{\rm ej,-4}^{-16/5}T_{\rm sd,3}^{6/5}t_{8}^{-6/5}+1$         &  $7.1\times10^{18}\xi^{8/5} M_{\rm ej,-4}^{-8/5}T_{\rm sd,3}^{3/5}t_{8}^{2/5}$     \\
\hline
          &  $t<T_{\rm{N1}}$                     &  $0.02\xi M_{\rm ej,-3}^{-1}T_{\rm sd,3}^{-1}t_{}+1$                           &  $7.8\times10^{9}\xi^{1/2} M_{\rm ej,-3}^{-1/2}T_{\rm sd,3}^{-1/2}t_{}^{3/2}$      \\
          &  $T_{\rm{N1}}<t<T_{\rm{sd}}$         &  $16.7\xi M_{\rm ej,-3}^{-1}T_{\rm sd,3}^{-1}t_{3}+1$                          &  $1.7\times10^{16}\xi^{2} M_{\rm ej,-3}^{-2}T_{\rm sd,3}^{-2}t_{3}^{3}$            \\
 Case III &  $T_{\rm{sd}}<t<T_{\rm{dec}}$        &  $16.7\xi M_{\rm ej,-3}^{-1}+1$                                                &  $1.7\times10^{17}\xi^{2} M_{\rm ej,-3}^{-2}t_{4}^{}$                              \\
          &  $T_{\rm{dec}}<t<T_{\rm N2}$         &  $3.5\xi^{1/8}n^{-1/8}t_{6}^{-3/8}+1$                                          &  $7.2\times10^{17}\xi^{1/4} n^{-1/4}t_{6}^{1/4}$                                   \\
          &  $t>T_{\rm{N2}}$                     &  $0.2\xi^{2/5}n^{-2/5}t_{8}^{-6/5}+1$                                          &  $2.8\times10^{18}\xi^{1/5} n^{-1/5}t_{8}^{2/5}$                                   \\
\hline
\end{tabular}
\end{table}

\clearpage

\begin{table}
\centering \caption{Temporal scaling indices of various parameters
in different temporal regimes for all dynamical cases.}
\begin{tabular}{lllllll}
\hline\hline

                                  &  $\gamma-1$     &  $R$            & $\nu_{\rm{a}}$         & $\nu_{\rm{m}}$ & $\nu_{\rm{c}}$ & $F_{\rm{\nu,max}}$ \\
\hline
Case I: $L_0 \sim 10^{47}~{\rm erg~s^{-1}},M_{\rm ej}\sim10^{-4}M_{\odot}$ \\
\hline
$t<T_{\rm{ma1}}$                  &  $1$            &  $\frac{3}{2}$  & $\frac{5p+4}{2(p+4)}$  & $\frac{5}{2}$  & $-\frac{7}{2}$ & $5$ \\
$T_{\rm{ma1}}<t<T_{\rm{N1}}$      &  $1$            &  $\frac{3}{2}$  & $\frac{1}{10}$         & $\frac{5}{2}$  & $-\frac{7}{2}$ & $5$\\
$T_{\rm{N1}}<t<T_{\rm{dec}}$      &  $1$            &  $3$            & $\frac{11}{5}$         & $4$            & $-6$           & $11$\\
$T_{\rm{dec}}<t<T_{\rm sd}$      &  $-\frac{1}{4}$ &  $\frac{1}{2}$  & $\frac{1}{5}$          & $-1$           & $-1$           & $1$ \\
$T_{\rm sd}<t<T_{\rm{ma2}} $      &  $-\frac{3}{8}$ &  $\frac{1}{4}$  & $0$                    & $-\frac{3}{2}$ & $-\frac{1}{2}$ & $0$ \\
$T_{\rm{ma2}}<t<T_{\rm{N2}}$      &  $-\frac{3}{8}$ &  $\frac{1}{4}$  & $-\frac{3p+2}{2(p+4)}$          & $-\frac{3}{2}$          & $-\frac{1}{2}$ & $0$\\
$t>T_{\rm{N2}}$                  &  $-\frac{6}{5}$ &  $\frac{2}{5}$  & $\frac{2-3p}{p+4}$     & $-3$           & $-\frac{1}{5}$ & $\frac{3}{5}$\\
\hline
Case II: $L_0 \sim 10^{49}~{\rm erg~s^{-1}},M_{\rm ej}\sim10^{-4}M_{\odot}$ \\
\hline
$t<T_{\rm{ma1}}$                  &  $1$            &  $\frac{3}{2}$  & $\frac{5p+4}{2(p+4)}$  & $\frac{5}{2}$  & $-\frac{7}{2}$ & $5$ \\
$T_{\rm{ma1}}<t<T_{\rm{N1}}$      &  $1$            &  $\frac{3}{2}$  & $\frac{1}{10}$         & $\frac{5}{2}$  & $-\frac{7}{2}$ & $5$\\
$T_{\rm{N1}}<t<T_{\rm sd}$       &  $1$            &  $3$            & $\frac{11}{5}$         & $4$            & $-6$           & $11$\\
$T_{\rm sd}<t<T_{\rm{ma2}}$      &  $-\frac{3}{8}$ &  $\frac{1}{4}$  & $0$                    & $-\frac{3}{2}$ & $-\frac{1}{2}$ & $0$ \\
$T_{\rm{ma2}}<t<T_{\rm{N2}}$      &  $-\frac{3}{8}$ &  $\frac{1}{4}$  & $-\frac{3p+2}{2(p+4)}$ & $-\frac{3}{2}$ & $-\frac{1}{2}$ & $0$\\
$t>T_{\rm{N2}}$                   &  $-\frac{6}{5}$ &  $\frac{2}{5}$  & $\frac{2-3p}{p+4}$     & $-3$           & $-\frac{1}{5}$ & $\frac{3}{5}$\\
\hline
Case III: $L_0 \sim 10^{49}~{\rm erg~s^{-1}},M_{\rm ej}\sim10^{-3}M_{\odot}$ \\
\hline
$t<T_{\rm{ma1}}$                  &  $1$            &  $\frac{3}{2}$  & $\frac{5p+4}{2(p+4)}$  & $\frac{5}{2}$  & $-\frac{7}{2}$ & $5$ \\
$T_{\rm{ma1}}<t<T_{\rm{N1}}$      &  $1$            &  $\frac{3}{2}$  & $\frac{1}{10}$         & $\frac{5}{2}$  & $-\frac{7}{2}$ & $5$\\
$T_{\rm{N1}}<t<T_{\rm sd}$       &  $1$            &  $3$            & $\frac{11}{5}$         & $4$            & $-6$           & $11$\\
$T_{\rm sd}<t<T_{\rm{dec}}$      &  $0$            &  $1$            & $\frac{3}{5}$          & $0$            & $-2$           & $3$\\
$T_{\rm{dec}}<t<T_{\rm{ma2}}$     &  $-\frac{3}{8}$ &  $\frac{1}{4}$  & $0$                    & $-\frac{3}{2}$ & $-\frac{1}{2}$ & $0$ \\
$T_{\rm{ma2}}<t<T_{\rm{N2}}$      &  $-\frac{3}{8}$ &  $\frac{1}{4}$  & $-\frac{3p+2}{2(p+4)}$ & $-\frac{3}{2}$ & $-\frac{1}{2}$ & $0$\\
$t>T_{\rm{N2}}$                   &  $-\frac{6}{5}$ &  $\frac{2}{5}$  & $\frac{2-3p}{p+4}$     & $-3$           & $-\frac{1}{5}$ & $\frac{3}{5}$\\
\hline
\end{tabular}
\end{table}

\clearpage

\clearpage

\clearpage


\begin{thebibliography}{37}
\expandafter\ifx\csname
natexlab\endcsname\relax\def\natexlab#1{#1}\fi

\bibitem[{{Abbott} {et~al.}(2009){Abbott}, {Abbott}, {Adhikari}, {Ajith},
  {Allen}, {Allen}, {Amin}, {Anderson}, {Anderson}, {Arain}, \& et~al.}]{ligo}
{Abbott}, B.~P., {Abbott}, R., {Adhikari}, R., {et~al.} 2009,
Reports on
  Progress in Physics, 72, 076901

\bibitem[{{Acernese} {et~al.}(2008){Acernese}, {Alshourbagy}, {Amico},
  {Antonucci}, {Aoudia}, {Astone}, {Avino}, {Baggio}, {Ballardin}, \&
  {Barone}}]{virgo}
{Acernese}, F., {Alshourbagy}, M., {Amico}, P., {et~al.} 2008,
Classical and
  Quantum Gravity, 25, 114045

\bibitem[{{Aloy} {et~al.}(2005){Aloy}, {Janka}, \& {M{\"u}ller}}]{aloy05}
{Aloy}, M.~A., {Janka}, H.-T., \& {M{\"u}ller}, E. 2005, \aap, 436,
273

\bibitem[{{Barthelmy} {et~al.}(2005){Barthelmy}, {Chincarini}, {Burrows},
  {Gehrels}, {Covino}, {Moretti}, {Romano}, {O'Brien}, {Sarazin},
  {Kouveliotou}, {Goad}, {Vaughan}, {Tagliaferri}, {Zhang}, {Antonelli},
  {Campana}, {Cummings}, {D'Avanzo}, {Davies}, {Giommi}, {Grupe}, {Kaneko},
  {Kennea}, {King}, {Kobayashi}, {Melandri}, {M\'esz\'aros}, {Nousek}, {Patel},
  {Sakamoto}, \& {Wijers}}]{barthelmy05a}
{Barthelmy}, S.~D., {Chincarini}, G., {Burrows}, D.~N., {et~al.}
2005, \nat,
  438, 994

\bibitem[{{Berger}(2011)}]{berger11}
{Berger}, E. 2011, \nat, 55, 1

\bibitem[{Bower} {et~al.}(2007)]{bower07}
 {Bower}, G.~C., {Destry}, S., {Bloom}, J.~S., et al. 2007, \apj, 666, 346

\bibitem[{Bower} \& {Saul}(2011)]{bower11}
 {Bower}, G.~C., \& {Saul}, D. 2011, \apj, 728, L14

\bibitem[Corsi
\& M{\'e}sz{\'a}ros(2009)]{corsi09} Corsi, A., \& M{\'e}sz{\'a}ros,
P.\ 2009, \apj, 702, 1171

\bibitem[{{Dai}(2004)}]{dai04}
{Dai}, Z.~G. 2004, \apj, 606, 1000

\bibitem[{{Dai} \& {Lu}(1998{\natexlab{a}})}]{dailu98b}
{Dai}, Z.~G., \& {Lu}, T. 1998{\natexlab{a}}, \aap, 333, L87

\bibitem[{{Dai} \& {Lu}(1998{\natexlab{b}})}]{dailu98a}
---. 1998{\natexlab{b}}, Physical Review Letters, 81, 4301

\bibitem[{{Dai} {et~al.}(2006){Dai}, {Wang}, {Wu}, \& {Zhang}}]{dai06}
{Dai}, Z.~G., {Wang}, X.~Y., {Wu}, X.~F., \& {Zhang}, B. 2006,
Science, 311,
  1127

\bibitem[{{Duncan} \& {Thompson}(1992)}]{duncan92}
{Duncan}, R.~C., \& {Thompson}, C. 1992, \apjl, 392, L9

\bibitem[{{Eichler} {et~al.}(1989){Eichler}, {Livio}, {Piran}, \&
  {Schramm}}]{eichler89}
{Eichler}, D., {Livio}, M., {Piran}, T., \& {Schramm}, D.~N. 1989,
\nat, 340,
  126

\bibitem[{{Fan} \& {Xu}(2006)}]{fanxu06}
{Fan}, Y.-Z., \& {Xu}, D. 2006, \mnras, 372, L19

\bibitem[Fan et al.(2013)]{fan13} Fan, Y.-Z., Wu, X.-F.,
\& Wei, D.-M.\ 2013, arXiv:1302.3328

\bibitem[{{Gao} \& {Fan}(2006)}]{gao06}
{Gao}, W.-H., \& {Fan}, Y.-Z. 2006, Chinese J. Astron. Astrophys.,
6, 513

\bibitem[{{Gehrels} {et~al.}(2005){Gehrels}, {Sarazin}, {O'Brien}, {Zhang},
  {Barbier}, {Barthelmy}, {Blustin}, {Burrows}, {Cannizzo}, {Cummings}, {Goad},
  {Holland}, {Hurkett}, {Kennea}, {Levan}, {Markwardt}, {Mason},
  {M\'esz\'aros}, {Page}, {Palmer}, {Rol}, {Sakamoto}, {Willingale},
  {Angelini}, {Beardmore}, {Boyd}, {Breeveld}, {Campana}, {Chester},
  {Chincarini}, {Cominsky}, {Cusumano}, {de Pasquale}, {Fenimore}, {Giommi},
  {Gronwall}, {Grupe}, {Hill}, {Hinshaw}, {Hjorth}, {Hullinger}, {Hurley},
  {Klose}, {Kobayashi}, {Kouveliotou}, {Krimm}, {Mangano}, {Marshall},
  {McGowan}, {Moretti}, {Mushotzky}, {Nakazawa}, {Norris}, {Nousek}, {Osborne},
  {Page}, {Parsons}, {Patel}, {Perri}, {Poole}, {Romano}, {Roming}, {Rosen},
  {Sato}, {Schady}, {Smale}, {Sollerman}, {Starling}, {Still}, {Suzuki},
  {Tagliaferri}, {Takahashi}, {Tashiro}, {Tueller}, {Wells}, {White}, \&
  {Wijers}}]{gehrels05}
{Gehrels}, N., {Sarazin}, C.~L., {O'Brien}, P.~T., {et~al.} 2005,
\nat, 437,
  851

\bibitem[{{Hotokezaka} {et~al.}(2012){Hotokezaka}, {Kiuchi}, {Kyutoku},
  {Okawa}, {Sekiguchi}, {Shibata}, \& {Taniguchi}}]{hotokezaka12}
{Hotokezaka}, K., {Kiuchi}, K., {Kyutoku}, K., {et~al.} 2012, ArXiv
e-prints (arXiv:1212.0905)

\bibitem[{{Klu{\'z}niak} \& {Ruderman}(1998)}]{kluzniak98}
{Klu{\'z}niak}, W., \& {Ruderman}, M. 1998, \apjl, 505, L113

\bibitem[{{Kramer} {et~al.}(2006){Kramer}, {Stairs}, {Manchester},
  {McLaughlin}, {Lyne}, {Ferdman}, {Burgay}, {Lorimer}, {Possenti}, {D'Amico},
  {Sarkissian}, {Hobbs}, {Reynolds}, {Freire}, \& {Camilo}}]{kramer06}
{Kramer}, M., {Stairs}, I.~H., {Manchester}, R.~N., {et~al.} 2006,
Science,
  314, 97

\bibitem[{{Kulkarni}(2005)}]{kulkarni05}
{Kulkarni}, S.~R. 2005, ArXiv Astrophysics e-prints
(arXiv:astro-ph/0510256)

\bibitem[{{Kuroda} {et~al.}(2010)}]{kuroda10}
Kuroda, K., \& LCGT Collaboration 2010, Classical and Quantum
Gravity, 27, 084004

\bibitem[Kyutoku et al.(2012)]{kyutoku12} Kyutoku, K., Ioka, K.,
\& Shibata, M. 2012, ArXiv Astrophysics e-prints (arXiv:1209.5747)

\bibitem[{{Li} \& {Paczy{\'n}ski}(1998)}]{lipaczynski98}
{Li}, L.-X., \& {Paczy{\'n}ski}, B. 1998, \apjl, 507, L59

\bibitem[{{Metzger} \& {Berger}(2012)}]{metzger12}
{Metzger}, B.~D., \& {Berger}, E. 2012, \apj, 746, 48

\bibitem[{{Metzger} {et~al.}(2011){Metzger}, {Giannios}, {Thompson},
  {Bucciantini}, \& {Quataert}}]{metzger11}
{Metzger}, B.~D., {Giannios}, D., {Thompson}, T.~A., {Bucciantini},
N., \&
  {Quataert}, E. 2011, \mnras, 413, 2031

\bibitem[{{Metzger} {et~al.}(2010){Metzger}, {Mart{\'{\i}}nez-Pinedo},
  {Darbha}, {Quataert}, {Arcones}, {Kasen}, {Thomas}, {Nugent}, {Panov}, \&
  {Zinner}}]{metzger10}
{Metzger}, B.~D., {Mart{\'{\i}}nez-Pinedo}, G., {Darbha}, S.,
{et~al.} 2010,
  \mnras, 406, 2650

\bibitem[{{Mimica} {et~al.}(2009){Mimica}, {Giannios}, \& {Aloy}}]{mimica09}
{Mimica}, P., {Giannios}, D., \& {Aloy}, M.~A. 2009, \aap, 494, 879

\bibitem[Morrison et al.(2004)]{morrison04} Morrison, I.~A.,
Baumgarte, T.~W., \& Shapiro, S.~L.\ 2004, \apj, 610, 941

\bibitem[{{Nakar} \& {Piran}(2011)}]{nakar11}
{Nakar}, E., \& {Piran}, T. 2011, \nat, 478, 82

\bibitem[{{Piran} {et~al.}(2012){Piran}, {Nakar}, \& {Rosswog}}]{piran12}
{Piran}, T., {Nakar}, E., \& {Rosswog}, S. 2012, ArXiv e-prints
(arXiv:1204.6242)

\bibitem[{{Rezzolla} {et~al.}(2011){Rezzolla}, {Giacomazzo}, {Baiotti},
  {Granot}, {Kouveliotou}, \& {Aloy}}]{rezzolla11}
{Rezzolla}, L., {Giacomazzo}, B., {Baiotti}, L., {et~al.} 2011,
\apjl, 732, L6

\bibitem[{{Rosswog} {et~al.}(2012){Rosswog}, {Piran}, \& {Nakar}}]{rosswog12}
{Rosswog}, S., {Piran}, T., \& {Nakar}, E. 2012, ArXiv e-prints
(arXiv:1204.6240)

\bibitem[{{Rosswog} {et~al.}(2003){Rosswog}, {Ramirez-Ruiz}, \&
  {Davies}}]{rosswog03}
{Rosswog}, S., {Ramirez-Ruiz}, E., \& {Davies}, M.~B. 2003, \mnras,
345, 1077

\bibitem[{{Rowlinson} \& {O'Brien}(2012)}]{rowlinson12}
{Rowlinson}, A., \& {O'Brien}, P. 2012, in -Ray Bursts 2012
Conference (GRB
  2012)

\bibitem[{{Rowlinson} {et~al.}(2010){Rowlinson}, {O'Brien}, {Tanvir}, {Zhang},
  {Evans}, {Lyons}, {Levan}, {Willingale}, {Page}, {Onal}, {Burrows},
  {Beardmore}, {Ukwatta}, {Berger}, {Hjorth}, {Fruchter}, {Tunnicliffe}, {Fox},
  \& {Cucchiara}}]{rowlinson10}
{Rowlinson}, A., {O'Brien}, P.~T., {Tanvir}, N.~R., {et~al.} 2010,
\mnras, 409,
  531

\bibitem[{{Sari} {et~al.}(1998){Sari}, {Piran}, \& {Narayan}}]{sari98}
{Sari}, R., {Piran}, T., \& {Narayan}, R. 1998, \apjl, 497, L17

\bibitem[{{Shibata} {et~al.}(2005){Shibata}, {Taniguchi}, \& {Ury{\=
  u}}}]{shibata05}
{Shibata}, M., {Taniguchi}, K., \& {Ury{\= u}}, K. 2005, \prd, 71,
084021

\bibitem[{{Taylor} \& {Weisberg}(1982)}]{taylor82}
{Taylor}, J.~H., \& {Weisberg}, J.~M. 1982, \apj, 253, 908

\bibitem[{{Zhang}(2013)}]{zhang13}
{Zhang}, B. 2013, \apjl, 763, L22

\bibitem[{{Zhang} \& {Kobayashi}(2005)}]{zhangkobayashi05}
{Zhang}, B., \& {Kobayashi}, S. 2005, \apj, 628, 315

\bibitem[{{Zhang} \& {M{\'e}sz{\'a}ros}(2001)}]{zhangmeszaros01a}
{Zhang}, B., \& {M{\'e}sz{\'a}ros}, P. 2001, \apjl, 552, L35

\bibitem[{{Zhang} \& {Yan}(2011)}]{zhangyan11}
{Zhang}, B., \& {Yan}, H. 2011, \apj, 726, 90

\end{thebibliography}
\end{document}